\documentclass[10pt,letterpaper,twocolumn]{article}
\usepackage[10pt,nocopyright]{sigmin}

\usepackage{listings}
\usepackage{color}
\usepackage{makecell}
\usepackage[square,comma,numbers,sort&compress]{natbib}
\usepackage[hyphens]{url}
\usepackage[breaklinks,colorlinks]{hyperref}
\usepackage[usenames,dvipsnames]{xcolor}
\hypersetup{citecolor=blue,linkcolor=blue}
\usepackage{amsmath,amsopn,amssymb,amsthm}
\usepackage{thmtools}
\usepackage[toc,page]{appendix}
\usepackage{subfigure}
\usepackage{endnotes,microtype,xspace,fancyvrb,multirow}
\usepackage{booktabs}
\usepackage{array,underscore,relsize}
\usepackage[T1]{fontenc}
\usepackage{inconsolata}
\usepackage{times}
\usepackage{fancyhdr,lastpage}
\usepackage{enumitem}
\usepackage[labelfont=bf,font=small,skip=1pt]{caption}
\usepackage[export]{adjustbox}
\usepackage{breakurl}
\usepackage{pifont}
\usepackage{tablefootnote}
\usepackage[linesnumbered,vlined,ruled,commentsnumbered]{algorithm2e}
\DontPrintSemicolon
\newcommand{\ra}[1]{\renewcommand{\arraystretch}{#1}}
\SetCommentSty{mycommfont}

\usepackage{amssymb}
\usepackage[normalem]{ulem}
\usepackage[hang,flushmargin]{footmisc}
\usepackage{textcomp}
\usepackage{graphicx}

\usepackage{epigraph}
\usepackage[compact]{titlesec}
\titleformat*{\section}{\large\bfseries}
\titleformat*{\subsection}{\normalsize\bfseries}
\titleformat*{\subsubsection}{\normalsize\bfseries}
\titlespacing*{\section}{0pt}{*2.5}{*1}
\titlespacing*{\subsection}{0pt}{*1.8}{*0.8}

\definecolor{mygreen}{rgb}{0,0.6,0}
\definecolor{mygray}{rgb}{0.5,0.5,0.5}
\definecolor{mymauve}{rgb}{0.58,0,0.82}

\hypersetup{
  colorlinks,
  linkcolor={red!50!black},
  citecolor={blue!50!black},
  urlcolor={blue!80!black}
}

\newcommand{\sys}{\textup{\mbox{MSched}}}

\newcommand{\stitleno}[1]{\vspace{.3ex}\noindent{\bf #1}}
\newcommand{\stitle}[1]{\vspace{1.ex}\noindent{\bf #1}}
\newcommand{\etitle}[1]{\vspace{0.7ex}\noindent{\em\underline{#1}}}

\newcommand{\squishlist}{
  \begin{list}{$\bullet$}{
    \setlength{\itemsep}{0pt}
    \setlength{\parsep}{3pt}
    \setlength{\topsep}{3pt}
    \setlength{\partopsep}{0pt}
    \setlength{\leftmargin}{1.5em}
    \setlength{\labelwidth}{1em}
    \setlength{\labelsep}{0.5em}
  }
}

\newcommand{\squishend}{
  \end{list}
}

\usepackage{tikz}

\usepackage{balance}
\usepackage{upgreek}

\begin{document}

\title{\Large \bf Towards Fully-fledged GPU Multitasking via Proactive Memory Scheduling}

\author{
    \\[1pt]
    Weihang Shen,
    Yinqiu Chen,
    Rong Chen\thanks{Corresponding author (\url{rongchen@sjtu.edu.cn}) and project leader.},
    Haibo Chen \\[4pt]
    \normalsize{Institute of Parallel and Distributed Systems, School of Computer Science, Shanghai Jiao Tong University} \\[-2pt]
}

\date{}
\maketitle

\frenchspacing

\begin{abstract}

Limited HBM capacity has become the primary bottleneck for
hosting a growing number of increasingly larger-scale GPU tasks.
While demand paging extends capacity via host DRAM,
it can incur order-of-magnitude slowdowns
due to the massive working sets of GPU workloads.
However, we observe that GPU memory accesses are inherently
predictable from kernel launch arguments, prior executions,
and the asynchronous execution nature.
Leveraging this opportunity, we propose {\sys},
an OS-level scheduler that extends GPU context switching
to include proactive working set preparation.
This approach coalesces fragmented, eventual, and expensive
page faults into a single efficient migration.
{\sys} combines parameterized templates with page fault feedback
to accurately predict working sets, and co-designs
the task scheduler and memory manager to resolve
global access order and enforce near-optimal page replacement.
Across scientific, deep learning, and LLM workloads,
{\sys} outperforms demand paging by up to 186.69$\times$
and whole-process swapping by up to 8.63$\times$.
{\sys} is open-source and has been integrated into
{\small{\url{https://github.com/XpuOS/nvxsched}}}.

\end{abstract}

\section{Introduction}
An increasing number of workloads today, especially AI-powered ones,
rely on GPU acceleration~\cite{multimedia-tcc20,clij-naturemethods20,
vectorsearch-fast25,gpudb-vldb23,aegaeon-sosp25}, creating a variety of scenarios
where multiple applications execute concurrently on a single
GPU~\cite{towards-arxiv25,xsched-osdi25,reef-osdi22,gpu-mt-survey-tpds22}.
For example, on personal devices like AI PCs, users may edit presentations
with GPU-powered text completion and image generation in the foreground,
while file indexing for RAG services runs in the background, sharing the sole GPU.
In locally deployed agent systems, where privacy and inference costs are key concerns,
multiple agents powered by different models discuss~\cite{crewai-github},
collaborate~\cite{anthropic-harness},
and execute subtasks concurrently~\cite{opencode-github,superpowers-parallel}.
In the cloud, model markets and serverless providers colocate cold
models receiving sporadic requests to increase deployment density
and reduce costs~\cite{aegaeon-sosp25,tgs-nsdi23}.

Prior research on GPU
multitasking~\cite{xsched-osdi25,reef-osdi22,gpreempt-atc25,effisha-ppopp17,
flep-asplos17,tally-asplos25}
has primarily focused on compute scheduling,
developing various techniques to multiplex GPU computation resources among tasks.
These systems implicitly assume that the GPU's high-bandwidth
memory (HBM){\footnote{\footnotesize{We use HBM to refer generically to on-device GPU memory.
}}}
can accommodate all concurrent applications.
However, this assumption no longer holds as a rapidly growing number of applications
incorporate AI~\cite{aiservice-apsys25,smartapp-arxiv23,aiui-arxiv24}
and rely on ever-larger models with substantial GPU memory demands.
Consequently, HBM capacity has become the primary bottleneck for hosting
an increasing number of larger-scale GPU applications concurrently.

The conventional wisdom for GPU memory oversubscription is
OS demand paging~\cite{tgs-nsdi23,nvshare-icse24} (e.g., Unified Memory~\cite{cuda-um}),
which transparently spills GPU memory pages to larger backing stores
such as host DRAM or SSDs~\cite{cuda-um,hip-um}.
During a GPU context switch,
only the minimal on-chip state (e.g., registers and shared memory) is swapped,
while the working set transfer is deferred.
Pages are swapped only when actually accessed,
using GPU virtual addressing and page faults to {passively} migrate
pages into HBM via interconnects like PCIe or NVLink C2C~\cite{nvlinkc2c,nvlinkc2c-benchmark}.

\begin{figure}[t]
	\vspace{-3mm}
	\begin{minipage}{.6\linewidth}
        \centering\includegraphics[width=\linewidth]{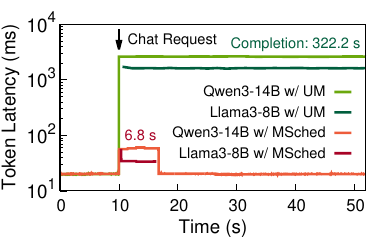}
    \end{minipage}
    \begin{minipage}{.39\linewidth}
        \centering\includegraphics[width=\linewidth]{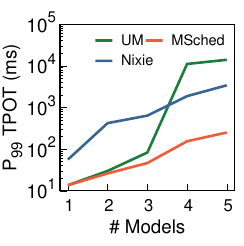}
    \end{minipage}
    \\[5pt]
    \begin{minipage}{1\linewidth}
        \caption{\emph{\small{(a) Token latency for a background agent task
        (Qwen3-14B, 8bit-quantized, 14.8\,GB) and an interactive chatbot task
        (Llama3-8B, 8bit-quantized, 8.5\,GB) using llama.cpp~\cite{llamacpp}
        on an NVIDIA RTX 5080 GPU (16\,GB),
        and (b) $P_{99}$ request TPOT when serving multiple independent
        Qwen3-32B (8bit) models using vLLM~\cite{vllm-github}
        on an NVIDIA H20 GPU (96\,GB),
        with demand paging (\textup{UM}), whole-process swapping
        (\textup{Nixie}~\cite{nixie-osdi26}), and {\sys},
        detailed in \S\ref{sec:eval:app-swap}.}}}
        \label{fig:intro-case}
    \end{minipage} \\[-15pt]
\end{figure}

This lazy mechanism works well for traditional CPU workloads,
which typically
have small working sets that can reside in physical memory,
so the occasional page faults impose acceptable overhead.
GPU programs, however, violate this premise: they often have
massive memory footprints, accessing several gigabytes of data
in just a few milliseconds~\cite{forest-isca25,earlyadaptor-ispass23,etc-asplos19}.
Moreover, GPU page faults are far more expensive than on CPUs. A fault locks
the TLB of the GPU compute unit (CU)~\cite{deepum-asplos23,cuda-um-opt},
stalling thousands of co-resident threads,
and traps into the CPU to swap pages and resolve the fault.
We measure this pathology in two concrete scenarios (see Fig.~\ref{fig:intro-case}):
On a user's PC, a Qwen3-14B coding agent continuously decodes in the background
while the user asks a Llama3-8B chatbot a short question
(100 input and 200 output tokens).
Enabling demand paging to manage oversubscribed memory causes a 113$\times$
slowdown on the background task and a 135$\times$ latency increase on the chatbot,
destroying the interactive experience.
For cloud providers, colocating multiple models with demand paging incurs
a 133$\times$ TPOT penalty once their working set exceeds HBM capacity
($>$\,3 models), severely violating SLOs.

\stitle{Opportunity.}
We observe that GPU applications have highly predictable memory access behaviors,
unlike CPU programs, which operate as black boxes to the OS.
Individual GPU kernels typically perform well-defined operations,
whose memory footprints can be inferred
from launch arguments or prior executions.
Moreover, GPU kernels are launched asynchronously in a straightforward sequence,
exposing their execution order to the OS beforehand.
This naturally provides an oracle for future GPU memory accesses
prior to execution.

\stitle{Key insight.}
This predictability motivates us to rethink the very definition
of context switching for GPU multitasking.
We argue that the notion of GPU ``context'' should expand beyond the minimal
on-chip execution state to include working set memory.
As GPU multitasking workloads often have working sets that exceed HBM capacity,
memory itself should be predictively ``scheduled''.
Rather than passively handling eventual and expensive page faults during execution,
the system should predict and prepare the
working set for the upcoming task
as an integral part of the context switch routine.

\stitle{Our approach.}
We introduce {\sys}, an OS-level scheduler designed for
GPU multitasking under memory oversubscription.
{\sys} extends GPU memory with host DRAM and treats working set memory
of a GPU task as a first-class citizen of its execution context.
{\sys} achieves this through two core mechanisms.
First, it predicts the memory regions each command will access
at launch time.
Second, {\sys} migrates the corresponding memory pages
of the upcoming task into GPU HBM during context switching.
This approach coalesces expensive, fragmented page faults into efficient,
batched memory transfers,
thereby significantly reducing the impact of memory oversubscription.

\stitle{Challenges.}
Realizing efficient memory scheduling is non-trivial.
The key challenge is ensuring the \textbf{prediction accuracy}
in both \underline{spatial} and \underline{temporal} aspects.
First, {\sys} must precisely identify the {spatial}
regions of memory each task will access within a given timeslice.
Under-prediction triggers costly page faults
that undermine proactive memory scheduling.
Conversely, the state-of-the-art system
(Nixie~\cite{nixie-osdi26} in Fig.~\ref{fig:intro-case}(b))
swaps entire process allocations, ensuring coverage but
incurring over-prediction that wastes bandwidth and drives TPOT
to seconds when serving multiple models, violating SLOs.
Second,
the {temporal} sequence of memory accesses determines
the optimal replacement policy, which is critical for efficiency.
However, in multitasking scenarios, context switches issued by different
scheduling policies cause memory accesses from concurrent tasks to
interleave in different ways,
making it difficult for the system to infer the
exact timing of each access and achieve optimal page replacement~\cite{opt-1966}.

\stitle{New techniques.}
To achieve \underline{spatial} accuracy, {\sys} employs
\emph{hybrid access prediction} that combines parameterized templates
and page fault feedback (\S\ref{sec:predict}).
For static accesses, whose target addresses are determined
before task execution, {\sys} models the mapping
from kernel arguments to memory regions using three templates
that capture common GPU access patterns.
For dynamic accesses, whose addresses are determined
during task execution, {\sys} uses statistical prediction
based on temporal continuity across kernel invocations,
learning from page fault feedback
and updating predictions through adaptive aging.
To achieve \underline{temporal} accuracy, {\sys} adopts
a \emph{scheduler-memory co-design} (\S\ref{sec:order}).
{\sys} exposes the task scheduling timeline---a common
scheduling primitive that describes future task order
and timeslice allocation---to the memory manager.
This enables the system to resolve interleaving and reconstruct
a global memory access sequence,
thereby enforcing a near-optimal replacement policy
aligned with the scheduler's decisions.

We evaluate {\sys} on NVIDIA RTX 5080 and H20 GPUs
under multitasking memory oversubscription.
Across scientific computing, deep learning, and LLM workloads,
{\sys} improves total throughput by 5.66--186.69$\times$
over demand paging (UM) and 1.37--8.63$\times$
over whole-process swapping (Nixie~\cite{nixie-osdi26}).
Under a priority-based policy,
{\sys} reduces the $P_{99}$ latency of real-time tasks by 4.06$\times$
over compute-only scheduling (XSched~\cite{xsched-osdi25}),
while boosting the throughput of colocated best-effort tasks by 2.44$\times$.
It further outperforms SUV~\cite{suv-micro24},
a single-task demand-paging optimization,
by 9.21$\times$ under heavy memory pressure.
For the scenarios in Fig.~\ref{fig:intro-case},
{\sys} cuts chatbot response time by 47.38$\times$ over UM
and improves $P_{99}$ TPOT by up to 16.09$\times$ over Nixie
when serving multiple colocated models.

\section{Background}

\subsection{Characterizing GPU Tasks}
\label{sec:bg:gpu-tasks}

Unlike general-purpose CPU workloads,
which often exhibit complex and branching control flows,
GPU tasks follow a sequential, host-driven execution model.
Typically, the host program allocates GPU memory through driver APIs for
all data, such as inputs, outputs, model parameters, and intermediates.
It then transfers input data to the GPU via memory copy commands.
Next, the host launches a series of GPU kernels to perform
computations on the data.
Finally, the results are copied back to the host.
From the OS perspective, a GPU task simply behaves as a sequence of
memory copy and GPU kernel commands~\cite{xsched-osdi25}.

A distinct feature of this model is the decoupling of command launch and execution.
Instead of executing immediately, commands launched by the host CPU are pushed into
a software-managed FIFO queue,
allowing the host thread to proceed to launch the next command.
The GPU's on-chip command processor,
e.g., NVIDIA GPU System Processor (GSP)~\cite{nvidia-gsp}, then fetches
and executes these commands from the queue in strict sequence.
This \textit{asynchronous} nature exposes a window of \textbf{future}
commands---queued but not yet executed---offering the OS a unique opportunity
to analyze and predict their resource requirements.

\subsection{Scheduling GPU Tasks}

To support multitasking, modern GPUs employ preemptive time-sharing scheduling
to multiplex hardware resources among concurrent GPU tasks
through context switching~\cite{nvidia-tsg,gcaps-ecrts24}.
This approach has become the mainstream paradigm in GPU design and deployment
due to its superior flexibility and robust fault and performance
isolation~\cite{effisha-ppopp17,flep-asplos17,reef-osdi22,tally-asplos25,gpreempt-atc25}.

\stitle{GPU context switching.}
\label{sec:gpu-context-switching}
Analogous to CPUs, a GPU execution context is traditionally defined as the
\emph{minimal architectural states} resident on the compute units (CUs)
that must be saved and restored to ensure correct execution continuation.
This includes register files, on-chip shared memory (scratchpad),
stack pointers, GPU page table bases, and control flow states
(e.g., program counters, active masks, barrier states),
with a total size of approximately 200\,KB per CU.
Modern GPUs, such as NVIDIA GPUs since the Pascal architecture and Intel Xe GPUs,
have natively supported preemption via context
switching~\cite{nvidia-cilp,xsched-osdi25,intel-gpu-specs,intel-gpu-context}
to schedule multiple tasks.
When preemption is requested (e.g., timeslice expiration),
the GPU command processor broadcasts an interrupt to the CUs,
immediately halting kernel execution and trapping into a software handler.
The handler saves the context to GPU HBM and terminates the current kernel.
The kernel of the subsequent task is then scheduled onto the CUs
to restore its context and resume execution,
thereby completing a typical GPU context switching operation.
This mechanism has been adopted by state-of-the-art GPU multitasking
systems~\cite{xsched-osdi25,gpreempt-atc25,gcaps-ecrts24} to enable flexible
and preemptive GPU scheduling.

\subsection{GPU Memory Multiplexing}

\stitleno{The untenable assumption of sufficient memory.}
While existing systems~\cite{xsched-osdi25,reef-osdi22,tally-asplos25,gpreempt-atc25}
have effectively scheduled GPU compute resources across multiple tasks,
they largely overlook the multiplexing of memory resources~\cite{towards-arxiv25}.
Most scheduling systems operate under an implicit assumption of sufficient memory:
they assume that the aggregate memory footprint of all concurrent tasks
can simultaneously reside in the GPU HBM.
Under this premise, context switching merely involves saving and restoring the
minimal architectural states (as described in \S\ref{sec:gpu-context-switching}),
enabling the upcoming task to execute seamlessly
as if it held exclusive access to the GPU.

However, this assumption is becoming increasingly untenable.
On one hand, the ubiquity of AI-powered applications---ranging from background
system agents to foreground interactive tools---has led to a surge
in the number of tasks contending for the
GPU~\cite{aiservice-apsys25,smartapp-arxiv23,aiui-arxiv24}.
On the other hand, the memory footprint of individual tasks is exploding
due to their massive model sizes and large caches~\cite{mooncake-fast25}.
Consequently, the total memory demand of concurrent tasks frequently exceeds
the HBM capacity of even high-end GPUs,
creating a barrier that simple compute scheduling cannot scale beyond.

\stitle{Demand paging as a workaround.}
\label{sec:bg:demand-paging}
To share the insufficient HBM across these memory-hungry tasks,
the conventional solution~\cite{tgs-nsdi23,nvshare-icse24}
is to enable demand paging for GPUs,
such as CUDA Unified Memory (UM)~\cite{cuda-um} and AMD HIP UM~\cite{hip-um},
which extends the GPU memory space with host DRAM
as a much larger and cheaper backing store.
In this architecture, the GPU driver maintains a unified virtual address space
where the physical pages are dynamically migrated between device HBM
and host DRAM.
When a running kernel accesses a page absent in HBM,
the GPU Memory Management Unit (MMU) raises a page fault,
stalls the faulting compute unit, and signals an interrupt to the CPU.
The GPU driver on the CPU catches the interrupt and migrates the faulting page
to HBM through high-bandwidth interconnects
like PCIe or NVLink C2C~\cite{nvlinkc2c,nvlinkc2c-benchmark}.
If the HBM is already full, the driver must evict pages to free up space.
Once the eviction and migration complete, the driver updates the GPU page table
and resumes the stalled compute unit.
By transparently and {passively} migrating pages upon access,
demand paging can be directly integrated into existing scheduling systems and
enable memory oversubscription.

\section{Rethinking GPU Context Switching}
\label{sec:think}

As shown in Fig.~\ref{fig:intro-case},
simply enabling demand paging (UM) for oversubscribed
concurrent GPU tasks collapses performance by up to 135$\times$.
This pathological degradation compels us to examine why demand
paging---effective in the CPU world---fails for GPU multitasking.
We identify three key limitations.

\etitle{1. Lengthy control plane of GPU page faults.}
Unlike CPUs, a GPU page fault is exorbitantly expensive due to hardware constraints.
First,
the GPU cannot handle page faults itself (\S\ref{sec:bg:demand-paging}).
It relies on the CPU-side driver
to resolve the mapping, issue multiple DMAs for data copy and page-table updates,
and synchronize with the GPU several times for correctness.
On an RTX 5080 GPU (PCIe 5.0), handling a single GPU page fault takes 31.79\,${\mu}s$,
of which only 1.35\,${\mu}s$ is data transfer---96\% goes to the control plane.
Second, a page fault triggered by a single GPU thread
locks the TLB of the entire CU and prevents further address translations
until the fault is resolved~\cite{deepum-asplos23,cuda-um-opt},
stalling thousands of concurrent GPU threads on that CU.
The massive parallelism of GPUs thus amplifies each stall into severe underutilization.

\begin{figure}[t]
    \centering
    \begin{minipage}{0.99\linewidth}
        \centering\includegraphics[width=\linewidth]{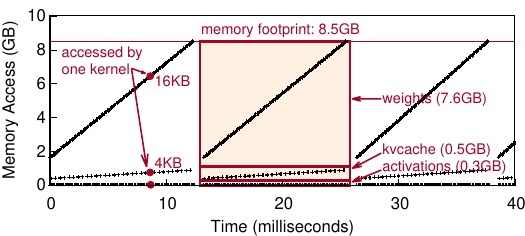}
    \end{minipage} \\[10pt]
    \begin{minipage}{1\linewidth}
        \caption{\emph{\small{GPU memory access pattern (sorted) of
        Llama3-8B (8bit-quantized) inference with llama.cpp
        on an NVIDIA RTX 5080 GPU.}}}
        \label{fig:access-pattern}
    \end{minipage} \\[-15pt]
\end{figure}

\etitle{2. Massive number of page faults.}
Beyond the cost of each individual fault, their sheer number is prohibitive.
Demand paging assumes workloads with small active working sets
that stay resident in physical memory.
GPU applications often violate this premise.
As illustrated in Fig.~\ref{fig:access-pattern}, a typical LLM
inference workload touches 8.5\,GB of data within just 12.7\,ms.
Serving such a working set through demand paging requires faulting
in an enormous number of pages within a very short window,
each paying the steep per-fault cost described above.

\etitle{3. Mutual page eviction across tasks.}
Worse yet, under multitasking, the aggregate working set of
concurrent tasks easily exceeds HBM capacity,
so keeping one task resident requires evicting another's pages.
Demand paging selects victims by recency,
unaware of which pages each task will need next.
The incoming task may therefore evict pages needed by
a soon-to-run task, forcing it to cold start
and fault them back in when it resumes.
This mutual eviction recurs every context switch,
and existing intra-task paging
optimizations~\cite{etc-asplos19,swapadvisor-asplos20,deepum-asplos23,suv-micro24}
cannot address it because they ignore inter-task interference.

\stitle{Opportunity.}
We observe that GPU workloads exhibit intrinsic predictability.
First, individual kernels typically perform well-defined computations,
making their memory accesses amenable to modeling.
For example, a matrix multiplication kernel accesses regions
determined by its matrix pointers and dimensions.
Second, each GPU task issues an explicit sequence of commands asynchronously,
exposing upcoming kernels and their per-task execution order.
These properties empower the OS to forecast the working set of GPU tasks.

\stitle{Predictive context switching.}
This predictability substantially changes GPU scheduling:
it largely reduces the memory-access uncertainty
that compels CPUs to rely on demand paging.
We therefore argue that GPU memory management should shift from passive
fault handling to predictive and proactive scheduling.
Specifically, we propose predictive context switching:
at each context switch, the system predicts the pages accessed
by the next task during its upcoming execution
and proactively loads them into HBM,
reclaiming space by evicting resident pages whose next use
is farthest in the future.
This paradigm addresses the three limitations above.
First, preparing pages before execution avoids costly fault handling
and the associated CU stalls on the critical path.
Second, batching working set migration into a single operation replaces
numerous individual faults with efficient bulk transfers
that better utilize interconnect bandwidth.
Third, since the future accesses of all tasks are predictable,
it evicts pages not used soon rather than another task's
imminently needed ones, avoiding mutual eviction.

\stitle{Challenges.}
To achieve efficient memory scheduling, the challenge lies in predicting
memory accesses {accurately} in both
spatial and temporal dimensions.

\etitle{Spatial accuracy.}
First, the system must precisely identify the pages each task will access
given the microsecond-scale time budget between command launch and execution.
Incomplete working set coverage (under-prediction) forces
the system to fall back to costly page faults.
On the other hand, simply preparing the entire process memory
ensures full coverage but incurs substantial over-prediction.
Transferring unused data wastes bandwidth and displaces useful pages,
increasing overall migration during subsequent execution.

\etitle{Temporal accuracy.}
Second, minimizing migration requires selecting eviction victims according to
the exact temporal order of future accesses across concurrent tasks.
Although each task executes commands in a fixed order,
multitask scheduling interleaves them, reshaping the global access sequence.
Eviction decisions must therefore align with the compute scheduling plan;
otherwise, the predicted order diverges from actual execution,
leading to suboptimal page replacement.

\section{{\sys} Overview}
\label{sec:overview}

\begin{figure}[t]
  \vspace{1mm}
  \begin{minipage}{1.\linewidth}
    \centering\includegraphics[width=\linewidth,viewport=93 365 304 509,clip=true]{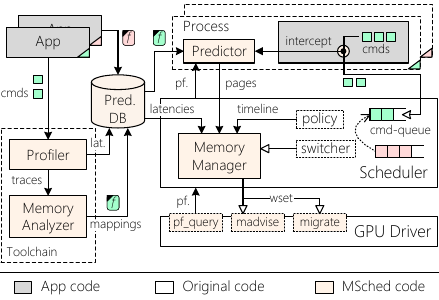}
  \end{minipage} \\[10pt]
  \begin{minipage}{1.\linewidth}
    \caption{\emph{\small{Architecture and workflow of {\sys}.}}}
    \label{fig:architecture}
  \end{minipage}  \\[-25pt]
\end{figure}

{\sys} realizes predictive context switching as an OS-level GPU memory scheduler,
pursuing three design goals:

\begin{itemize}[leftmargin=2.3ex,itemsep=-0.5ex,topsep=0.25ex]

\item \textbf{Efficiency.}
{\sys} minimizes the context-switch overhead of GPU memory multiplexing
to sustain near-ideal performance under oversubscription.

\item \textbf{Transparency.}
{\sys} requires no application modifications or recompilation
and remains fully compatible with closed-source stacks such as CUDA.

\item \textbf{Extensibility.}
{\sys} adopts a modular design that extends existing GPU scheduling
components through clean interfaces, supporting diverse scheduling policies.

\end{itemize}

\stitle{Architecture.}
Fig.~\ref{fig:architecture} highlights the components introduced by {\sys}:
an offline toolchain and prediction database, a runtime predictor and
memory manager, and extensions to the existing scheduler and driver.

\etitle{Prediction database and toolchain.}
A system-wide database stores per-kernel prediction mappings
from launch arguments to accessed memory regions,
along with execution latencies.
An offline toolchain automatically derives these mappings through
profiling and analysis,
enabling the runtime to predict both memory accesses and execution duration.

\etitle{Runtime predictor.}
A predictor library preloaded into each application intercepts GPU commands
and evaluates the mappings using actual launch arguments.
It supplements the predicted pages with driver-reported page faults
to capture dynamic accesses missed by the mappings.

\etitle{Memory manager.}
At each context switch, the memory manager combines the task timeline,
per-command predictions, and execution latencies
to determine which pages to retain or evict
and prepare the incoming working set.
The timeline provides a policy-agnostic interface,
keeping memory management decoupled from scheduling policy.

\etitle{Extended scheduler and driver.}
{\sys} augments the existing scheduler to preemptively schedule GPU tasks
and expose the task timeline, and extends the GPU driver with interfaces
to enforce the eviction order, migrate pages between HBM and host DRAM,
and report page fault feedback.

\stitle{Workflow.}
The offline toolchain profiles kernels and stores
prediction mappings and execution latencies in the database.
Developers can also register mappings for more precise prediction.
A predictor loaded to each application predicts memory pages
accessed by each intercepted GPU command.
At each context switch, the scheduler passes its task timeline
to the memory manager, which combines it with predictions and latencies
to determine the incoming working set and page eviction order.
The manager then directs the GPU driver to evict selected pages
and populate the incoming working set.

\section{Memory Access Prediction}
\label{sec:predict}

Memory scheduling requires preparing the working set
of the upcoming task.
A recent work proposes \textbf{allocation-based switching},
which migrates the entire allocated memory of the incoming process~\cite{nixie-osdi26}.
Although this approach covers the working set,
it transfers substantial amounts of data unused within the timeslice,
e.g., reserved but inactive KV cache,
wasting bandwidth and evicting reusable pages.
Efficient memory scheduling therefore requires accurate working set
prediction, minimizing both false negatives (under-prediction)
and false positives (over-prediction).
Prediction must also be lightweight,
as the asynchronous time window between command launch and execution
can be as short as a few microseconds~\cite{picker-eurosys26},
leaving little budget for per-kernel analysis.

A GPU task comprises memory copy and compute kernel commands.
The memory footprint of a copy command is explicitly specified
in its API arguments as source, destination, and transfer size.
Compute kernels derive access addresses by performing computation
or transformation on certain original variables such as pointers,
dimensions, and indices.
We refer to these address-determining data as
\textit{initial variables} $x$
and model the memory access behavior of the kernel as
an address expression $addrs=f(x)$.
We classify accesses as \textbf{static}
if all initial variables exist before task execution,
and \textbf{dynamic} if any are created during task execution,
such as expert IDs generated by a preceding gating kernel
to select expert weights in MoE LLMs.
For static accesses, {\sys} identifies variables $x$ among kernel arguments
and models the mapping $f$ using parameterized templates.
For more challenging dynamic accesses, {\sys} uses a statistical
approach based on page fault feedback to predict the accessed regions.

\subsection{Predicting Static Memory Accesses}
\label{sec:predict:static}

Static accesses can be predicted by identifying the initial
variables $x$ and modeling the mapping $f$.
We observe that these variables are typically supplied
as kernel launch arguments, such as data pointers,
sizes, and shapes~\cite{picker-eurosys26}.
For example, \texttt{vecadd(A,B,C,N)} adds vectors at pointer
\texttt{A} and \texttt{B} to \texttt{C},
each containing \texttt{N} elements.
All its initial variables (\texttt{A}, \texttt{B}, \texttt{C}, and \texttt{N})
are thus provided as kernel arguments.
We therefore explore how to identify $x$ among launch arguments
and model $f$ to map their values to accessed regions.

A recent approach~\cite{phos-sosp25} is to match kernel
arguments against live buffers.
It tracks GPU memory allocation and deallocation requests
to maintain the address range of each buffer.
If an argument points into a known buffer,
the system conservatively predicts that the kernel will access
the entire buffer.
While conceptually simple, buffer boundaries are too coarse
in practice.
Modern DL frameworks such as PyTorch~\cite{pytorch-malloc,towards-arxiv25},
TensorFlow~\cite{tensorflow-malloc}, and JAX~\cite{jax-malloc} pre-allocate
large memory pools and internally sub-allocate them to tensors.
Similarly, LLM serving systems~\cite{llamacpp,vllm-github} reserve GBs of
monolithic buffers for the entire model weights, activations, and KV cache,
then slice them for individual layers or requests.
Consequently, this \textbf{buffer-based} approach
\textbf{over-predicts} a kernel's working set
from a few tensors or a single layer's data
(e.g., about 20\,KB in Fig.~\ref{fig:access-pattern})
to nearly the entire process memory,
similar to the allocation-based switching.

\begin{figure}[t]
	\begin{minipage}{.98\linewidth}
        \centering\includegraphics[width=\linewidth]{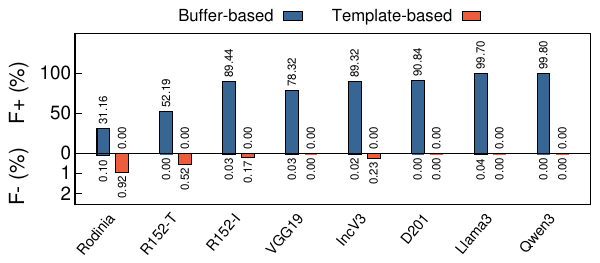}
    \end{minipage} \\[8pt]
    \begin{minipage}{1\linewidth}
        \caption{\emph{\small{Kernel-level false negative rate \textup{(F-)}
		and false positive rate \textup{(F+)} of two prediction approaches.
        Workloads: \textup{Rodinia} scientific computing
        benchmark~\cite{rodinia-iiswc09},
        ResNet152 {\underline{T}}raining using PyTorch,
        ResNet152, VGG19, InceptionV3, DenseNet201
        {\underline{I}}nference using PyTorch,
        Llama3-8B and Qwen3-8B inference using llama.cpp.}}}
        \label{fig:prediction:false}
    \end{minipage} \\[-15pt]
\end{figure}

Our empirical analysis in Fig.~\ref{fig:prediction:false}
confirms that this approach is inadequate.
While the false positive rate is 31\% for scientific computing benchmarks,
it surges to an average of 80\% for standard deep learning workloads.
Most critically, for LLM inference,
the false positive rate reaches an alarming 99.8\%,
bringing GBs of unnecessary migration.

\begin{figure}[t]
	\begin{minipage}{.49\linewidth}
        \centering\includegraphics[width=\linewidth]{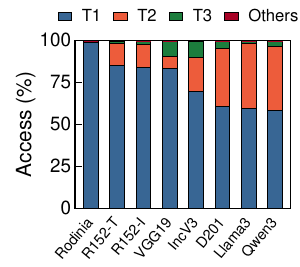}
    \end{minipage}
    \begin{minipage}{.49\linewidth}
        \centering\includegraphics[width=\linewidth]{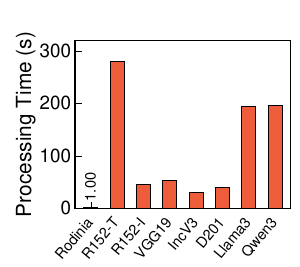}
    \end{minipage} \\[8pt]
    \begin{minipage}{1\linewidth}
        \caption{\emph{\small{(a) Share of memory accesses per template and
        (b) offline processing time of the toolchain
        across GPU workloads in Fig.~\ref{fig:prediction:false}.}}}
        \label{fig:prediction:type-offline}
    \end{minipage} \\[-15pt]
\end{figure}

\stitle{Template-based prediction.}
Despite the flexibility of GPU programming,
we observe that individual kernels have well-defined semantics
and often exhibit regular access patterns.
Our analysis of memory traces from representative GPU workloads
spanning scientific computing, DL inference and training, and LLMs
identifies three basic addressing patterns, whose prevalence is
summarized in Fig.~\ref{fig:prediction:type-offline}\,(a).
{\sys} captures these patterns using three built-in
parameterized templates.
Each template describes memory regions with a \textit{base address}
specified by a particular pointer argument and a \textit{layout}
(i.e., size and shape) defined as follows:

\begin{itemize}[leftmargin=2.3ex,itemsep=-0.5ex,topsep=0.25ex]

\item \textbf{T1: Fixed-size access} ($\sim$75\%).
The accessed region has a fixed size across all kernel invocations,
either independent of launch arguments or determined by invariant arguments.
This pattern dominates because kernels are often invoked in a consistent
manner throughout the application's lifecycle,
with most launch arguments unchanged.

\item \textbf{T2: Linear-size access} ($\sim$20\%).
The accessed region is contiguous,
but its size scales linearly with the product of specific launch arguments.
This pattern is common when arguments specify element counts
or tensor dimensions in machine learning.
For example, in kernel \texttt{vecadd(A,\,B,\,C,\,N)}, the accessed sizes of
\texttt{A}, \texttt{B}, and \texttt{C} scale linearly with element
count \texttt{N}, while in \mbox{\texttt{matmul(A,\,B,\,C,\,M,\,N,\,K)}},
they scale with matrix dimensions \texttt{M$\times$K}, \texttt{K$\times$N},
and \texttt{M$\times$N}, respectively.

\item \textbf{T3: Strided access} ($\sim$5\%).
The access layout consists of multiple
discontiguous memory chunks separated by strides, where the stride,
chunk size, and chunk count scale linearly with the product of specific launch arguments.
This pattern typically arises in ML workloads,
when kernels operate on specific dimensions of high-dimensional tensors.

\end{itemize}

\noindent
A kernel's static memory access behavior ($f$) can be modeled
as the combination of multiple argument-to-region expressions,
each following one of the access templates.
{\sys} also allows developers to add new templates
to support diverse access patterns.
The per-kernel prediction mappings $f$ are maintained
in a system-wide database.
GPU developers can explicitly specify these mappings when implementing
kernels to precisely describe their accessed regions, and register them
in the database during application or library installation.
{\sys} also provides an \textit{automated toolchain}
that derives prediction mappings through
memory access profiling and offline analysis.
To demonstrate its effectiveness, all prediction mappings used in our
evaluation are generated automatically without manual effort.

\stitle{Profiler.}
We developed an offline profiler based on the CUPTI~\cite{cupti-nvidia}
profiling tool and NVBit~\cite{nvbit-micro19},
an instrumentation tool for GPU kernel binaries,
with native support for all kernels including JIT-compiled, fused,
and CUDA Graph kernels.
The profiler first runs the application with the original kernels to measure
the execution latency of each command.
Since GPU commands have deterministic latencies~\cite{clockwork-osdi20,rammer-osdi20,reef-osdi22},
these measurements are used at runtime to predict which commands will execute
in the next timeslice (\S\ref{sec:order}).
It then performs a second run that intercepts all kernels,
captures their launch arguments, and instruments every memory access instruction
to trace the touched memory addresses.

\stitle{Memory analyzer.}
Per-kernel prediction mappings are inferred from the collected memory traces
and launch arguments using an offline analyzer.
It first identifies candidate base-pointer arguments by matching
64-bit argument values to the starting addresses of traced regions.
Next, it attempts to match the \textit{layout} of these regions
against the three templates in order.
If the region size remains constant across profiled invocations,
the analyzer selects T1.
Otherwise, it enumerates 64- and 32-bit integer arguments and
their products to identify linear proportionality with
the region size (T2) or stride (T3).
For C-style struct arguments, the analyzer slices them into
64- and 32-bit integers and treats them as independent candidates.
Each successful match produces a prediction expression with
the template type, argument indices and offsets, and linear coefficients.
The offline overhead is acceptable: profiling and analysis together
take 48.77\,s on average, shown in Fig.~\ref{fig:prediction:type-offline}\,(b).

\stitle{Online prediction.}
The online predictor is implemented as a preloaded library.
It intercepts all GPU commands, retrieves prediction mappings,
and evaluates them to calculate the precise memory access regions
of each command (within a microsecond).
The regions are then aligned to page boundaries and
attached to command metadata for proactive scheduling.

\stitle{Prediction accuracy.}
\label{sec:predict:accuracy}
Across the scientific computing, DL, and dense LLM workloads
in Fig.~\ref{fig:prediction:false}, template-based prediction achieves
near-perfect coverage (an average false-negative rate of 0.23\%)
with zero false positives.
The remaining misses stem primarily from dynamic accesses,
which we address next using fault-driven prediction.

\subsection{Predicting Dynamic Memory Access}
\label{sec:predict:dynamic}

For dynamic accesses, evaluating the address expression requires
initial variables generated as intermediate results during execution.
The accessed regions therefore cannot be resolved from template-based mappings alone.
For example, in MoE LLM inference,
a gating kernel outputs activated expert IDs,
which subsequent kernels read to access the corresponding expert weights.
Such dynamic accesses account for 32.16\% of the memory footprint
in Qwen3-30B-A3B inference.
Waiting for these variables leaves insufficient time for memory preparation.
Moreover, they are typically passed through GPU-resident
data structures rather than returned to the CPU,
such as expert ID buffers and block tables used by
PagedAttention~\cite{pagedattention-sosp23} to locate KV cache blocks.
Obtaining and interpreting such variables would require
additional synchronization and knowledge of their data structures.

To predict dynamic accesses in advance,
{\sys} adopts a statistical approach.
We observe that dynamic memory accesses often exhibit
temporal continuity across successive kernel invocations.
Recent studies~\cite{moe-predict-isca26,klotski-asplos25,lina-atc23}
report skewed expert activations and similar expert selections
between adjacent executions.
In PagedAttention, kernels reuse the KV cache prefix
from the previous step and access newly appended KV entries.
These observations motivate predicting dynamic memory accesses
from per-kernel access history, without waiting for
the underlying variables to be generated.

\stitle{Fault-based access tracking.}
Learning this history requires visibility into
dynamic accesses during execution.
However, GPU page tables lack CPU-style accessed bits,
while instrumenting and tracking every GPU memory access instruction
would incur substantial runtime overhead.
To overcome this challenge,
{\sys} repurposes GPU page faults as a lightweight feedback signal
to distinguish dynamic accesses and learn per-kernel access history.
We modify the kernel-mode driver to record page fault addresses
and timestamps, and use CUPTI~\cite{cupti-nvidia} to obtain the begin
and end times of each kernel.
By correlating these records, {\sys} identifies the kernel
responsible for each page fault.
These faults reveal the dynamically accessed pages,
which were missed by template-based prediction.
At subsequent launches, {\sys} includes these pages
in its predictions for proactive migration.

\stitle{Adaptive prediction aging.}
Page faults provide only partial feedback:
once a page is proactively migrated, the absence of further faults
does not indicate whether it is accessed again.
{\sys} therefore assigns each fault-learned page a prediction lifetime
to keep predictions up to date.
Each page starts with a lifetime of one timeslice unit,
which decreases by one each time the task executes a timeslice (aging).
Once the lifetime expires, the page is excluded from predictions.
Each time the page faults again, {\sys} doubles its previously
assigned prediction lifetime.
Repeated faults extend the prediction, while aging expires
it without renewed access evidence.

The two prediction techniques in {\sys} are complementary.
Template-based prediction precisely captures static accesses
and immediately reflects changes in kernel arguments.
Fault-driven prediction extends coverage to complex
dynamic accesses and supplements template predictions
by learning from observed misses.
Together, they enable accurate working set prediction
for memory scheduling.

\section{Access Order Prediction}
\label{sec:order}

Beyond predicting accessed memory regions,
efficient memory scheduling further demands \underline{temporal accuracy},
as access order determines page replacement
under Belady's optimal algorithm~\cite{opt-1966}.
The asynchronous and sequential nature of GPU tasks,
combined with our per-command memory prediction,
offers a unique opportunity to apply the in-theory policy,
which is elusive for CPUs due to opaque execution flows.

However, exploiting this opportunity under multitasking is non-trivial,
as context switching fundamentally alters the global order of command execution.
Different scheduling policies introduce diverse interleavings of commands from
concurrent tasks, disrupting the sequential continuity within individual tasks.
Therefore, without awareness of the scheduling plan,
it is impossible to accurately predict the global order of future memory accesses,
thereby undermining the implementation of the optimal replacement policy (OPT~\cite{opt-1966}).

\subsection{Predicting the Global Page Access Order}
\label{sec:globle-order}

\stitle{Task scheduling timeline as the Rosetta Stone.}
To address this challenge, {\sys} co-designs the task scheduler and memory manager.
We extended the scheduling policy module to expose its task scheduling timeline
to the memory manager at each context switch.
The task scheduling timeline is an ordered sequence of task entries,
recording currently executing commands and allocated timeslices for each task,
akin to the run queue in OS schedulers~\cite{modern-os,linux-runqueue}.
This simple structure provides the ground truth for future execution---which
task will execute, for how long, and in what order---and
effectively decouples memory management from scheduling
with a clean interface, allowing {\sys} to support diverse policies.

\begin{figure}[t]
  \begin{minipage}{1.\linewidth}
    \centering\includegraphics[width=\linewidth,viewport=125 72 404 183,clip=true]{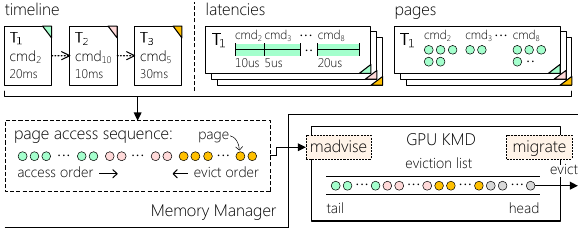}
  \end{minipage} \\[10pt]
  \begin{minipage}{1.\linewidth}
    \caption{\emph{\small{An example of memory scheduling in {\sys}.
	T$_{1}$, T$_{2}$, and T$_{3}$ are scheduled in turn.
	The memory manager orders pages in the eviction list via \texttt{madvise}
	based on the page access sequence.}}}
    \label{fig:mm}
  \end{minipage}  \\[-15pt]
\end{figure}

\stitle{Global access order.}
The task timeline, the per-command execution latencies,
and the page predictions from \S\ref{sec:predict}
jointly define a global access order.
This order can be derived conceptually in three steps.
(1) Per-task command sequence.
Within each task, commands execute in program order.
The command at which each task will resume is read from the timeline,
and its process-local command queue is traversed forward from that point.
(2) Mapping commands onto timeslices.
The profiled latencies of successive commands are accumulated
to determine how many commands fit within each allocated timeslice
and identify the command at which the timeslice expires.
A command spanning a timeslice boundary is replicated in both timeslices.
(3) Interleaving across tasks.
Following the timeslice order in the timeline,
these per-task command segments are concatenated into a global execution
sequence across all concurrent tasks.
Associating each command with its predicted pages establishes
the global page access sequence shown in Fig.~\ref{fig:mm}.

\subsection{Scheduling Memory with the Predicted Order}
\label{sec:predict-order}

\stitle{From access order to eviction order.}
Armed with the global page access order,
{\sys} decides which pages to retain and which to evict.
Since replacement costs are uniform across pages,
Belady's OPT is to evict the page whose next access
is farthest in the future~\cite{opt-1966}.
Pages are therefore ordered for eviction
by decreasing predicted next-use time.

{\sys} constructs the eviction order by traversing the page access sequence
\emph{in reverse}, moving each referenced page to the tail of the eviction list.
Fig.~\ref{fig:mm} illustrates this for three tasks scheduled in turn.
From head to tail, the resulting list contains pages unreferenced
across the timeline (grey),
T$_{3}$'s working set for its 30\,ms timeslice (orange),
T$_{2}$'s 10\,ms working set (pink),
and T$_{1}$'s 20\,ms working set (cyan, cmd$_{2}$ to cmd$_{8}$).
Pages needed farthest in the future are evicted first,
while the immediate working set of the next task is protected at the tail.
For pages referenced multiple times, the earliest upcoming reference
determines their final position.

\stitle{Memory scheduling decision.}
{\sys} applies this order at context switch to schedule GPU memory.
It populates the incoming working set according to the page access sequence
before the task resumes, evicting pages from the list head
when HBM space is insufficient.
\S\ref{sec:impl} details how these decisions
are enforced through the GPU driver.

\stitle{Updating the eviction order.}
\label{sec:order:dynamic}
This eviction order is not fixed.
First, the task timeline may change due to scheduling decisions
or the arrival of new tasks.
Second, workloads such as model training and iterative LLM decoding
launch GPU commands intermittently,
causing the predicted page sets to change over time.
{\sys} therefore recomputes the eviction order at every context switch.
The overhead is negligible,
remaining below 1\,ms for 16 concurrent tasks (\S\ref{sec:eval:overhead}).
In practice, this per-switch update keeps the eviction order close to optimal,
as confirmed by results on LLM decoding, training,
and workloads with dynamically arriving tasks (\S\ref{sec:eval}).

\section{Implementation and Optimization}
\label{sec:impl}

\stitleno{Memory manager.}
{\sys} implements memory scheduling through a distributed memory manager.

\etitle{Distributed metadata.}
The memory manager comprises a central coordinator daemon
and a helper library preloaded into each application process.
The helper records each GPU command in a process-local queue,
along with its predicted working set (pages)
and offline-profiled execution latency.
This keeps fine-grained metadata local to application processes,
reducing communication overhead with the coordinator.

\etitle{Enforcing the eviction order.}
To construct the eviction order,
the coordinator traverses the timeline in reverse
and instructs each helper via shared-memory IPC
to process the accesses within the corresponding timeslice.
Each IPC specifies the timeslice's offset
from the task's current execution position.
The helper identifies the corresponding command segment and traverses
its predicted accesses in reverse order (\S\ref{sec:predict-order}),
calling \texttt{madvise} to move the referenced pages
to the tail of the eviction list in the kernel-mode driver (KMD).

\stitle{GPU driver extensions.}
{\sys} extends the GPU KMD with interfaces that support
proactive memory scheduling and optimizes
its page migration.\footnote{\footnotesize{All modifications are made to
the open-source NVIDIA GPU kernel modules
and are portable to other GPU vendors such as AMD and Intel.}}

\etitle{Driver interfaces.}
{\sys} adds \texttt{ioctl}s to the KMD to expose its internal
least-recently-faulted (LRF) eviction list of HBM-resident pages.
Newly faulted pages join the tail,
while eviction proceeds from the head.
The \texttt{madvise} interface lets userspace relocate
specific pages to the tail to enforce the eviction order.
The \texttt{migrate} interface evicts pages in the list to host DRAM,
then populates the requested pages into HBM.
\texttt{pf_query} returns fault-migrated pages and their timestamps
as feedback for dynamic access prediction (\S\ref{sec:predict:dynamic}).

\etitle{Pipelined migration.}
Since page eviction and population dominate context-switch latency,
maximizing migration throughput is critical.
Modern GPUs expose multiple Copy Engines (CEs) for concurrent
DMA~\cite{cuda-guide,amd-ce,demystify-rtas24},
while full-duplex interconnects such as PCIe and
NVLink-C2C~\cite{GH200-study-arxiv24} allow device-to-host eviction
and host-to-device population to proceed in parallel.
However, current GPU drivers~\cite{nvidia-driver-580} leave this
parallelism untapped,
serializing four DMA operations for each page swap:
unmapping the victim page, evicting it to the host (D2H),
populating the target page (H2D), and establishing its mapping.
{\sys} instead pipelines these operations across CEs.
It initiates unmapping and D2H eviction on CE0, then launches
H2D population and mapping on CE1 once sufficient space is reclaimed,
while CE0 proceeds to the next victim.
This tight pipeline overlaps eviction with population,
effectively saturating the full-duplex interconnect.

\stitle{Task scheduler extensions.}
{\sys} extends XSched~\cite{xsched-osdi25} with two lightweight hooks
to prepare memory and expose scheduling information,
maintaining a decoupled design.

\etitle{Context-switch hook.}
After suspending a task, the scheduler invokes
the memory manager to evict pages and populate the incoming working set
before resuming execution.

\etitle{Policy support.}
Scheduling policies commonly maintain internal structures
similar to the task timeline.
{\sys} exposes this information to the memory manager through
an \texttt{update\_timeline} interface.
With fewer than ten lines of code, {\sys} supports diverse
user-defined policies such as priority,
earliest-deadline-first, and least-laxity-first
(e.g., as provided by XSched~\cite{xsched-osdi25}).
Our evaluation primarily uses the round-robin (RR) policy---natively
adopted by many GPUs~\cite{amd-rr,intel-gpu-specs,nvidia-tsg}
for fairness and simplicity---to show the advantages of memory scheduling.
We also evaluate a priority policy (Fig.~\ref{fig:eval-xsched})
that reduces the latency of randomly arriving real-time tasks,
demonstrating extensibility.

\section{Evaluation}
\label{sec:eval}

\begin{figure}[t]
	\begin{minipage}{.325\linewidth}
        \centering\includegraphics[width=\linewidth]{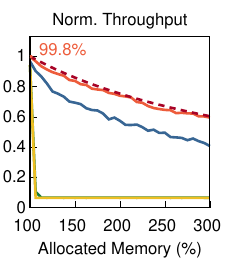}
    \end{minipage}
    \begin{minipage}{.325\linewidth}
        \centering\includegraphics[width=\linewidth]{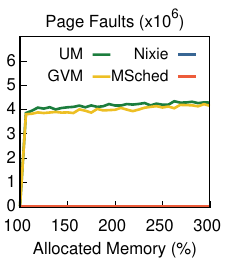}
    \end{minipage}
    \begin{minipage}{.325\linewidth}
        \centering\includegraphics[width=\linewidth]{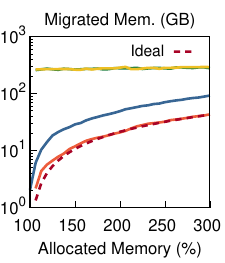}
    \end{minipage} \\[8pt]
    \begin{minipage}{1\linewidth}
        \caption{\emph{\small{Comparison of (a) end-to-end throughput,
        (b) page fault count and (c) memory migration volume
        required to complete computation equivalent to one second of in-HBM execution,
        among \textup{UM}, \textup{GVM~\cite{gvm-paper}},
        \textup{Nixie~\cite{nixie-osdi26}}, {\sys},
        and theoretical limit (\textup{Ideal}).}}}
        \label{fig:eval-micro}
    \end{minipage} \\[-10pt]
\end{figure}

\stitleno{Experimental setup.}
The evaluation is mainly conducted on two testbeds:
a workstation equipped with an Intel Core Ultra 9 285K CPU,
96\,GB of DDR5 memory, and an NVIDIA RTX 5080 GPU (16\,GB);
and a server with Intel Xeon Platinum 8458P CPUs,
1\,TB of DDR5 memory, and an NVIDIA H20 GPU (96\,GB of HBM).
Both GPUs connect to the host via PCIe 5.0$\times$16.
We primarily use the round-robin policy
to match the default time-sharing behavior of commodity GPUs.
Note that {\sys} operates at the OS level and is transparent
to applications, enabling proactive memory scheduling
without any code modification or recompilation.

\stitle{Baselines.}
We mainly compare {\sys} against UM~\cite{cuda-um},
GVM~\cite{gvm-paper,gvm-github}, Nixie~\cite{nixie-osdi26},
and an Ideal baseline.
For UM, we intercept GPU memory allocation and deallocation calls
to transparently use CUDA Unified Memory for demand paging.
GVM provides per-application memory and page-fault isolation.
We split HBM capacity equally among applications
via the GVM KMD.
Nixie is a state-of-the-art system that multiplexes the GPU
through whole-process swapping.
It adopts allocation-based switching
and selects the first started application as the eviction victim.
Ideal is the theoretical limit under perfect working-set prediction,
the OPT replacement policy,
and the measured full-duplex PCIe bandwidth.

\subsection{Mechanistic Effectiveness}
\label{sec:eval:micro}

\stitleno{Methodology.}
We first evaluate the mechanistic advantages
of accurate proactive memory scheduling.
We handcraft two types of representative GPU tasks:
compute-intensive matrix multiplication tasks and
memory-bound multi-head attention tasks.
For each type, we launch two processes (four total) that
continuously issue tasks.
The aggregate memory footprint is increased
by adjusting the number of matrices or the sequence length.
Fig.~\ref{fig:eval-micro} reports the end-to-end throughput
(normalized to exclusive in-HBM execution without oversubscription),
the page fault count and migration volume required to complete
computation equivalent to one second of in-HBM execution.

\stitle{Performance breakdown.}
Due to the large working sets of GPU workloads,
UM experiences severe page thrashing,
producing approximately four million page faults and 300\,GB of migration,
with a 14.4$\times$ slowdown under oversubscription.
Although GVM isolates memory usage and page faults across applications,
its per-application quotas remain insufficient to accommodate
large working sets, causing intra-application thrashing
and performance similar to UM.
Nixie avoids memory contention and page faults through
whole-process swapping, but its coarse migration granularity
and naive replacement policy incur substantial unnecessary movement.
In contrast, {\sys} accurately identifies per-kernel working sets
and predicts access timing from scheduling information
to guide OPT-based page replacement.
This reduces migration volume by 2.10--2.82$\times$ relative to Nixie
and improves throughput by 1.46$\times$ at 300\% memory allocation.
Across memory pressures, {\sys} closely tracks \textit{Ideal}
in throughput and migration volume, demonstrating its effectiveness.
Without oversubscription (100\%), {\sys} retains 99.8\% of throughput,
incurring negligible overhead (0.2\%).

\subsection{End-to-End Application Performance}
\label{sec:eval:app}

\begin{table}[t]
    \vspace{2mm}
    \centering
    \renewcommand\cellgape{\Gape[1.5pt]}
    \begin{minipage}{1.\linewidth}
        \caption{\textit{\small{GPU application combinations description.
		}}}
        \label{tab:app-workload}
    \end{minipage}
    \begin{minipage}{1.\linewidth}
    \ra{1.02}
    \centering
    \small{
    \begin{tabular}{@{~}l l l l@{~}}
        \toprule
        Comb. & Type && Task \\
        \midrule
        A & Scientific && kmeans, srad, cfd, nn \\
        B & DNN        && {R}Net, {V}GG, {I}nception, {D}Net \\
        C & Hybrid DL  && R, V, I, D, Llama3-8B (llama.cpp) \\
        D & LLM        && Llama3-8B, Qwen3-8B (llama.cpp) \\
        E & MoE LLM    && Qwen3-30B-A3B (llama.cpp) \\
        F & LLM Serve  && Qwen3-32B (vLLM, H20 testbed)\\
        \bottomrule
    \end{tabular}
    }
    \end{minipage} \\[-10pt]
\end{table}

\begin{figure*}[t]
    \centering
    \begin{minipage}{.36289\linewidth}
        \centering\includegraphics[width=\linewidth]{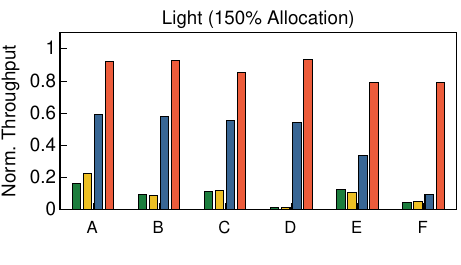}
    \end{minipage}%
    \begin{minipage}{.31527\linewidth}
        \centering\includegraphics[width=\linewidth]{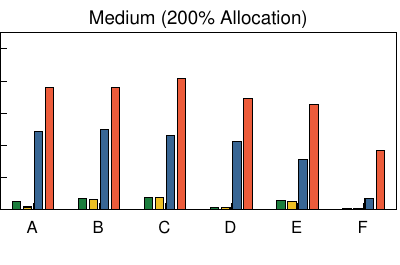}
    \end{minipage}%
    \begin{minipage}{.32184\linewidth}
        \centering\includegraphics[width=\linewidth]{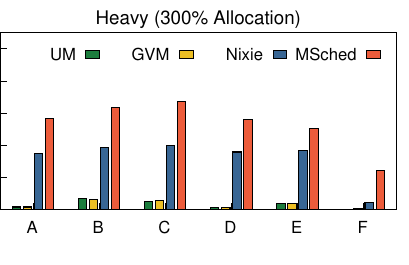}
    \end{minipage} \\[-3pt]
    \begin{minipage}{1\linewidth}
        \caption{\emph{\small{Comparison of normalized task throughput
        among \textup{UM}, \textup{GVM~\cite{gvm-paper,gvm-github}},
        \textup{Nixie~\cite{nixie-osdi26}},
        and {\sys} across different workload combinations
        under (a) Light, (b) Medium, and (c) Heavy memory pressure.}}}
        \label{fig:eval-app}
    \end{minipage} \\[-15pt]
\end{figure*}

\stitleno{Methodology.}
We evaluate {\sys} using six combinations of
real applications (Table~\ref{tab:app-workload}).
\textbf{A (Scientific)}: four tasks from Rodinia~\cite{rodinia-iiswc09}
benchmark suite---K-means clustering (kmeans), image denoising (srad),
fluid dynamics solver (cfd), and nearest-neighbors search (nn).
\textbf{B (DNN)}: PyTorch inference with ResNet152 (RNet),
VGG19, InceptionV3, and DenseNet201 (DNet).
\textbf{C (Hybrid DL)}: all tasks in B alongside
8bit-quantized Llama3-8B inference using llama.cpp.
\textbf{D (LLM)}: concurrent llama.cpp instances
of 8bit-quantized Llama3-8B and Qwen3-8B,
representing an on-device multi-agent workload.
\textbf{E (MoE LLM)}: multiple 2bit-quantized
Qwen3-30B-A3B llama.cpp instances.
\textbf{F (LLM Serve)}: multiple vLLM serving instances
of 8bit-quantized Qwen3-32B,
simulating cloud model markets~\cite{aegaeon-sosp25}.
Combination F runs on the H20 testbed, while all others
run on the RTX 5080 testbed.
Each task runs in a closed-loop as an independent process
under three oversubscription levels:
\textbf{\underline{L}ight} (total memory allocation = 150\% of HBM capacity),
\textbf{\underline{M}edium} (200\%), and \textbf{\underline{H}eavy} (300\%),
controlled by scaling problem sizes (A), batch sizes (B, C),
or model instance counts (D--F),
with memory balanced evenly across tasks.
We report task throughput normalized to in-HBM execution
in Fig.~\ref{fig:eval-app}.

\stitle{Performance.}
For combinations A--D, which are dominated by static accesses,
{\sys} achieves speedups of 5.66--65.67$\times$ over UM,
4.04--64.31$\times$ over GVM, and 1.53--1.78$\times$ over Nixie.
The benefits also extend to workloads with dynamic accesses:
{\sys} outperforms Nixie by 1.37--2.36$\times$ for MoE inference (E).
On larger GPUs (F with H20), {\sys} achieves speedups of
18.36--186.69$\times$ over UM and 5.41--8.63$\times$ over Nixie.
The larger gains for F underscore the benefits of precise
working set prediction and OPT-guided migration on GPUs
with larger HBM capacities and workload footprints,
where interconnect bandwidth becomes a more pronounced bottleneck.
These results align with the findings in
Fig.~\ref{fig:eval-micro},
confirming {\sys}'s effectiveness and advantages.

\subsection{Performance Impact of Prediction Accuracy}

\begin{figure}[t]
    \centering
    \includegraphics[width=\linewidth]{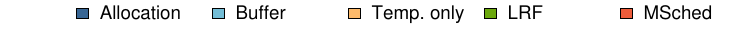}\par\nointerlineskip
    \begin{minipage}{.44\linewidth}
        \centering\includegraphics[width=\linewidth]{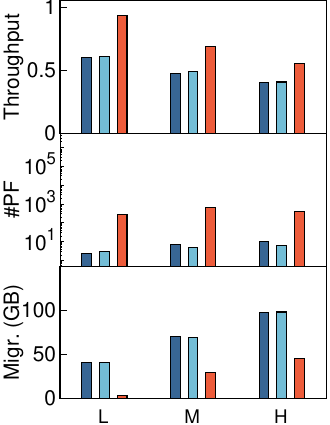}
    \end{minipage}%
    \begin{minipage}{.33\linewidth}
        \centering\includegraphics[width=\linewidth]{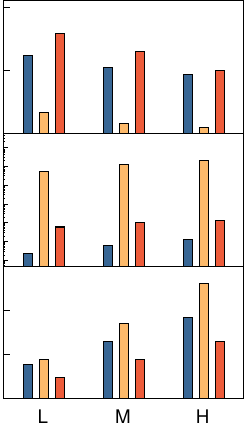}
    \end{minipage}%
    \begin{minipage}{.23\linewidth}
        \centering\includegraphics[width=\linewidth]{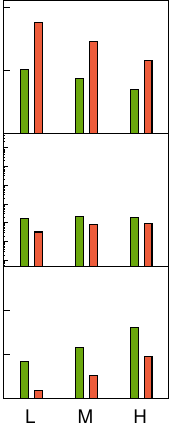}
    \end{minipage} \\[8pt]
    \begin{minipage}{1\linewidth}
        \caption{\emph{\small{Normalized throughput, page fault count,
        and migration volume of memory scheduling under
        Light (L), Medium (M), and Heavy (H) memory pressure
        (a) using allocation-based switching, buffer-based prediction,
        and \textup{\sys} on Comb.~D (dense LLMs),
        (b) using allocation-based switching, template-only prediction,
        and \textup{\sys} on Comb.~E (MoE LLMs),
        and (c) using LRF and \textup{\sys}'s OPT
        page replacement policy on Comb.~A--E.}}}
        \label{fig:eval:ablation}
    \end{minipage} \\[-10pt]
\end{figure}

\stitleno{Template-based static access prediction.}
As shown in Fig.~\ref{fig:prediction:false} (\S\ref{sec:predict:accuracy}),
our template-based prediction significantly outperforms
the buffer-based approach in false positive rate.
To quantify the impact of this precision gap,
we run Comb.~D, a dense LLM workload
dominated by static accesses, with each prediction technique
and allocation-based switching.
We measure normalized throughput, page fault counts, and migration volume.
As shown in Fig.~\ref{fig:eval:ablation}\,(a),
all three approaches incur sporadic page faults.
However, buffer-based prediction inflates the prediction
to nearly the entire process allocation, resulting in migration volume
and throughput comparable to allocation-based switching.
By avoiding over-prediction, {\sys}---here driven solely by
template-based prediction---reduces migration volume by up to 11.67$\times$
and improves throughput by up to 1.53$\times$
over buffer-based prediction.

\stitle{Fault-driven dynamic access prediction.}
We compare allocation-based switching, template-only prediction,
and {\sys} using Comb.~E.
Template-based prediction alone cannot resolve dynamic accesses
driven by expert selection in MoE.
Gaps in working-set coverage incur noticeable page faults
and cause page replacement to deviate from OPT,
increasing migration and reducing throughput.
Allocation-based switching covers these accesses,
but at the cost of over-prediction,
increasing migration and degrading performance.
As shown in Fig.~\ref{fig:eval:ablation}\,(b),
fault-driven prediction enables {\sys} to reduce page faults
by 99.89--99.92\% over template-only prediction
and improve throughput by 4.78$\times$, 8.62$\times$, and 10.82$\times$
under Light, Medium, and Heavy memory pressure, respectively.
For Comb.~F (vLLM), disabling fault-driven prediction
has no measurable performance impact (results omitted).
This is because the PagedAttention kernels are surrounded by
KV cache management and movement kernels with
template-predictable static accesses.
Since {\sys} migrates the union of predicted working sets
within a timeslice, these KV blocks are already covered
by template-based prediction.
This demonstrates the robustness of our prediction scheme,
which preserves working-set coverage even when predictions
for individual kernels are incomplete.

\stitle{OPT-based page replacement.}
We compare the default least-recently-faulted (LRF) policy
built into the GPU driver with OPT-based policy ({\sys}) using
Comb.~A--E in Table~\ref{tab:app-workload} on the RTX 5080 testbed.
The LRF policy is unaware of future kernel execution and memory accesses,
and may evict pages needed by upcoming kernels,
causing unnecessary migration.
Fig.~\ref{fig:eval:ablation}\,(c) reports the normalized throughput,
page faults, and migration volume,
geometrically averaged across combinations.
The OPT-based policy reduces migration by
4.69$\times$, 2.20$\times$, and 1.69$\times$
and improves end-to-end throughput by
1.74$\times$, 1.69$\times$, and 1.65$\times$
under Light, Medium, and Heavy memory pressure, respectively.
These gains come from evicting pages with the most distant predicted use,
preserving limited HBM capacity for imminent computation.

\begin{figure}[t]
	\begin{minipage}{.60\linewidth}
        \centering\includegraphics[width=\linewidth]{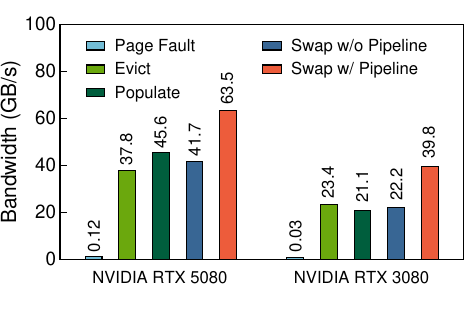}
    \end{minipage}
    \begin{minipage}{.39\linewidth}
        \centering\includegraphics[width=\linewidth]{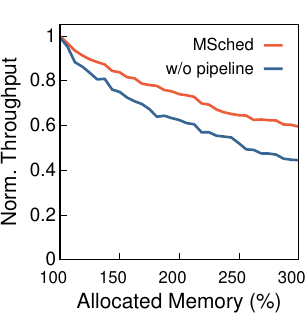}
    \end{minipage} \\[8pt]
    \begin{minipage}{1\linewidth}
        \caption{\emph{\small{(a) Page migration bandwidth using different methods
        on two different platforms, and (b) comparison of end-to-end throughput
        with (w/) and without (w/o) pipelined migration.}}}
        \label{fig:eval-seq}
    \end{minipage} \\[-10pt]
\end{figure}

\subsection{Accelerating Page Migration}

We further analyze the impact of the page migration bandwidth
between GPU HBM and host DRAM on memory scheduling
and evaluate our pipelined migration optimization.
For generality, we introduce a \textbf{new testbed} equipped with
Intel Core i7-13700, 64\,GB DDR4, and an \textbf{RTX 3080} GPU
(10\,GB, PCIe 4.0$\times$16).
We first measure the bandwidth of page eviction (D2H with unmapping),
population (H2D with mapping), fault-driven swapping (evict then populate),
and proactive swapping with and without pipelining.
As shown in Fig.~\ref{fig:eval-seq}\,(a),
the effective swap bandwidth without pipelining is bounded by the average
of D2H and H2D---41.7\,GB/s on RTX 5080 and 22.2\,GB/s on RTX 3080.
Our pipelined approach overlaps D2H and H2D
to exploit the full-duplex interconnect,
boosting the effective swap bandwidth to 63.5\,GB/s (1.52$\times$)
on RTX 5080 and 39.8\,GB/s (1.79$\times$) on RTX 3080.
Note that the throughput is still below the PCIe 5.0$\times$16
bandwidth (64\,GB/s$\times$2) due to a hardware bottleneck
in Intel Arrow Lake desktop CPUs~\cite{arrowlake-bottleneck}.

We then assess the end-to-end impact
using the workload from \S\ref{sec:eval:micro} on RTX 5080.
In Fig.~\ref{fig:eval-seq}\,(b), the improved bandwidth yields
a 1.34$\times$ end-to-end speedup at 300\% subscription.
These results highlight the criticality of migration bandwidth for proactive scheduling.
As interconnects continue to scale, with PCIe bandwidth doubling per generation
(PCIe 7.0$\times$16 at 256\,GB/s$\times$2),
and new fabrics such as CXL~\cite{cxl-survey-csur24} and NVLink C2C
(450\,GB/s$\times$2)~\cite{nvlinkc2c,nvlinkc2c-benchmark,GH200-study-arxiv24}
becoming available, {\sys} will benefit proportionally.

\begin{figure}[t]
	\hspace{1mm}
	\begin{minipage}{.61\linewidth}
        \centering\includegraphics[width=\linewidth]{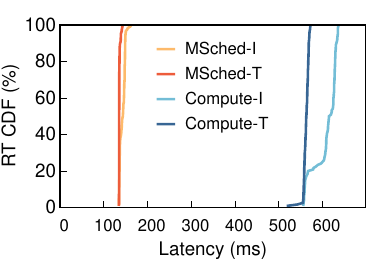}
    \end{minipage}
	\hspace{2mm}
    \begin{minipage}{.31\linewidth}
        \centering\includegraphics[width=\linewidth]{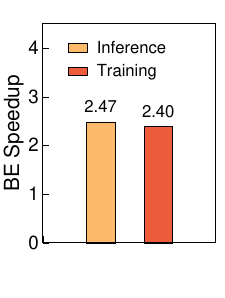}
    \end{minipage}\\[8pt]
    \begin{minipage}{1\linewidth}
        \caption{\emph{\small{(a) Latency CDF of RT tasks
        and (b) throughput speedup of BE tasks
        for {\sys} compared with compute-only scheduling
        (\textup{XSched~\cite{xsched-osdi25}}) on RTX~5080
        using a priority scheduling policy
        when inference (I) or training (T) serves as the BE task.}}}
        \label{fig:eval-xsched}
    \end{minipage} \\[-10pt]
\end{figure}

\subsection{Extensibility to a Priority Policy}

To evaluate the policy-agnostic design of {\sys} beyond round-robin,
we use a \textbf{priority policy} and compare performance
with and without memory scheduling.
Both configurations use the same compute scheduler,
while the latter relies on demand paging under memory oversubscription.
We emulate a cloud colocation scenario mixing
dynamically arriving Real-Time (RT) tasks---subject to
strict SLOs and running exclusively at high priority---with
Best-Effort (BE) tasks that improve utilization at low priority
and are preempted upon RT arrivals (1\,req/s Poisson).
We use ResNet152 inference as the RT task
and either ResNet152 inference (I) or training (T) as the BE task,
forming two test cases.
Batch sizes are tuned so that RT uses 6\,GB and BE uses 12\,GB.

Fig.~\ref{fig:eval-xsched}\,(a) shows that with memory scheduling,
{\sys} reduces the $P_{99}$ latency of RT tasks by 4.06$\times$
on average, improving service quality.
Meanwhile, its scheduling-aware replacement policy
reduces data movement and boosts BE task throughput by 2.44$\times$
on average (Fig.~\ref{fig:eval-xsched}\,(b)), increasing overall GPU utilization.
These results demonstrate that the same timeline interface
effectively supports priority scheduling under workloads
with dynamic task arrivals.

\subsection{Overhead Analysis}
\label{sec:eval:overhead}

\begin{figure}[t]
	\begin{minipage}{.5\linewidth}
        \centering\includegraphics[width=\linewidth]{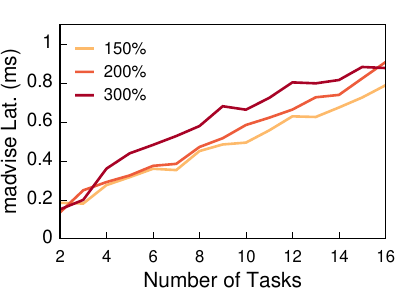}
    \end{minipage}
    \begin{minipage}{.49\linewidth}
        \centering\includegraphics[width=\linewidth]{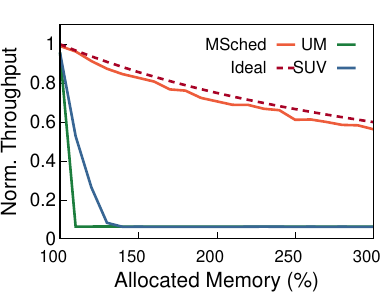}
    \end{minipage}\\[8pt]
    \begin{minipage}{1\linewidth}
        \caption{\emph{\small{(a) Control-plane overhead (\texttt{madvise})
        of {\sys} under different task counts and memory allocations.
        (b) End-to-end throughput of \textup{UM}, \textup{SUV~\cite{suv-micro24}},
        {\sys}, and \textup{Ideal} under different memory allocations on NVIDIA RTX 3080.}}}
        \label{fig:eval:overhead-suv}
    \end{minipage} \\[-10pt]
\end{figure}

Beyond the data-plane cost of page migration,
{\sys} introduces control-plane overhead,
primarily incurred by the synchronous \texttt{madvise} calls
each task issues to reorder the driver's eviction list
during context switching.
Using the workload from \S\ref{sec:eval:micro},
we measure the aggregate \texttt{madvise} latency (including IPC) during a single
context switch under varying task counts and memory allocations.
As shown in Fig.~\ref{fig:eval:overhead-suv}\,(a),
the latency scales linearly with the number of tasks,
i.e., the number of \texttt{madvise} calls.
We also observe a slight latency increase under higher memory pressure,
due to slower page table lookups within the driver
as more pages are allocated.
Even so, within typical task-count ranges (tens of tasks or fewer),
the overhead remains under 1\,ms,
which is negligible compared to the data migration latency.

\subsection{Comparison with Intra-Task Paging Optimization}

Existing OS-level optimizations for GPU demand paging
target single-task execution. In Fig.~\ref{fig:eval:overhead-suv}\,(b),
we use the workloads in \S\ref{sec:eval:micro} to compare {\sys}
against SUV~\cite{suv-micro24}, a state-of-the-art system
that employs compile-time static analysis to identify data hotness
and guide memory replacement and prefetching.
Because SUV relies on data structures in legacy GPU drivers
that are incompatible with the latest GPUs,
we conduct this comparison on our RTX 3080 testbed.
Notably,
SUV requires access to kernel source code,
precluding support for common deep learning frameworks like PyTorch and llama.cpp
that use closed-source kernel libraries (e.g., cuDNN, cuBLAS).

\begin{figure}[t]
	\vspace{-3mm}
	\begin{minipage}{.49\linewidth}
        \centering\includegraphics[width=\linewidth]{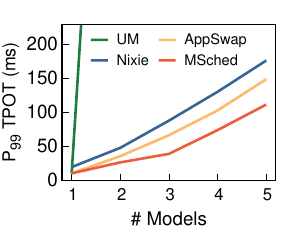}
    \end{minipage}
	\begin{minipage}{.49\linewidth}
        \centering\includegraphics[width=\linewidth]{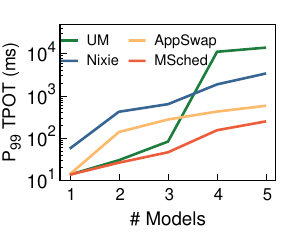}
    \end{minipage}\\[5pt]
    \begin{minipage}{1\linewidth}
        \caption{\emph{\small{$P_{99}$ request TPOT for serving multiple
        concurrent vLLM model instances of
        (a) 8bit-quantized Llama3-8B on RTX 5080 GPU
        and (b) 8bit-quantized Qwen3-32B on H20 GPU,
        using \textup{UM}, \textup{Nixie~\cite{nixie-osdi26}},
        vLLM's built-in swapping (AppSwap), and {\sys}.}}}
        \label{fig:eval-vllm}
    \end{minipage} \\[-15pt]
\end{figure}

SUV performs poorly in multitasking workloads,
while {\sys} delivers substantial gains,
achieving a 9.21$\times$ throughput
improvement over SUV at 300\% memory allocation.
This gap exposes a fundamental limitation of single-task optimizations:
they are oblivious to the drastic working set transitions
induced by context switching.
Without scheduling-aware coordination, per-task policies collide,
causing migration conflicts and thrashing.
{\sys} avoids this by inferring the cross-task
memory access order from the task timeline
and enforcing a scheduling-aligned page replacement policy.

\subsection{Comparison with Application-Level Swapping}
\label{sec:eval:app-swap}

Extending the multi-model serving scenario in Fig.~\ref{fig:intro-case}(b),
we use vLLM to serve multiple independent instances of Llama3-8B
on RTX 5080 and Qwen3-32B on H20, emulating a model market workload,
where providers colocate cold models receiving sporadic requests
to increase deployment density
while meeting QoS requirements~\cite{aegaeon-sosp25}.
We compare {\sys} against UM, Nixie, and AppSwap,
an application-level approach that uses vLLM's built-in
\texttt{sleep} and \texttt{wake_up} APIs~\cite{vllm-sleep}
to suspend models and swap them to host DRAM.
We use prompts and output lengths from the ShareGPT
dataset~\cite{sharegpt-huggingface} and generate requests
for each model with a Poisson process at 0.1\,RPS.
For UM, Nixie, and {\sys}, each model runs in a separate vLLM process.
For AppSwap, we rotate among multiple models within a single vLLM process.

Fig.~\ref{fig:eval-vllm} shows that with oversubscribed allocation
(2--5 models), {\sys} reduces $P_{99}$ TPOT by up to 2.25$\times$
and 1.70$\times$ on RTX 5080, and 16.09$\times$ and 5.95$\times$ on H20,
compared to Nixie and AppSwap, respectively.
{\sys} avoids migrating unused allocations, including reserved
KV-cache regions, by accurately predicting the working set.
UM outperforms Nixie and AppSwap on H20 when active working sets
can fit in HBM ($\leqslant$\,3 models),
since Nixie and AppSwap still perform whole-model swaps.
AppSwap benefits from in-process scheduling and richer application
semantics, outperforming Nixie's cross-process approach.
Overall, {\sys} improves service quality for colocated models
and enables higher deployment density.

\section{Related Work}

\stitleno{GPU scheduling systems.}
Prior research has extensively explored GPU compute multiplexing techniques.
EffiSha~\cite{effisha-ppopp17}, FLEP~\cite{flep-asplos17},
PipeSwitch~\cite{pipeswitch-osdi20}, REEF~\cite{reef-osdi22},
XSched~\cite{xsched-osdi25},
and others~\cite{chimera-asplos15,gcaps-ecrts24,tally-asplos25,gpreempt-atc25}
enable low-latency task switching through GPU preemption.
Meanwhile, approaches such as NVIDIA MPS~\cite{nvidia-mps}, MIG~\cite{nvidia-mig},
libsmctrl~\cite{libsmctrl-rtas23}, Baymax~\cite{baymax-asplos16},
Paella~\cite{paella-sosp23}, Orion~\cite{orion-eurosys24},
SGDRC~\cite{sgdrc-ppopp25}, and LithOS~\cite{lithos-sosp25}
improve utilization through spatial sharing and resource isolation.
Clockwork~\cite{clockwork-osdi20}, TGS~\cite{tgs-nsdi23},
and Shepherd~\cite{shepherd-nsdi23} develop scheduling policies
for deep learning workloads.
However, these compute-centric works either implicitly assume that the GPU HBM
can accommodate the aggregate memory footprint of all concurrent tasks,
or delegate memory oversubscription to the native demand paging mechanism
which brings severe performance degradation.
GVM~\cite{gvm-paper} enforces GPU memory capacity limits for applications.
Nixie~\cite{nixie-osdi26} is a state-of-the-art scheduling system
that swaps entire processes
and selects the first started application for eviction.
In contrast, {\sys} treats working set memory as a first-class
citizen of the GPU context, precisely identifying and proactively
scheduling the memory with a near-optimal replacement policy.

\stitle{Demand paging optimizations for a single GPU task.}
Prior work optimized GPU demand paging (typically CUDA UM) under
memory oversubscription by exploiting program access characteristics
to guide page eviction and prefetching.
Examples include SUV~\cite{suv-micro24}, DeepUM~\cite{deepum-asplos23},
Sentinel~\cite{sentinel-hpca21}, SwapAdvisor~\cite{swapadvisor-asplos20},
and others~\cite{snurhac-hpdc21,huvm-atc22,earlyadaptor-ispass23}.
Others, such as Forest~\cite{forest-isca25}, ETC~\cite{etc-asplos19},
and SMC~\cite{smc-icpp24}, redesign hardware to improve the demand paging routine
and introduce specialized units (e.g., memory compressors)
to reduce page migration overhead.
However, these systems are tailored for single-task execution and unaware of concurrency.
Applying them in multitasking scenarios creates severe
cross-task page conflicts and thrashing,
as evidenced in Fig.~\ref{fig:eval:overhead-suv}\,(b).
In comparison, {\sys} co-designs the scheduler and memory manager,
aligning page replacement with scheduling policies.

\stitle{In-application GPU memory swapping.}
Prior work implemented numerous application-specific swapping techniques
to circumvent HBM constraints, offloading model
parameters~\cite{clockwork-osdi20,pipeswitch-osdi20,moelightning-asplos25,
sirius-atc25,powerinfer-sosp24}, KV cache~\cite{lmcache-arxiv25},
and tensors~\cite{capuchin-asplos20} to host DRAM.
Others spill GPU data to
disk~\cite{powerinfer2-arxiv24,llminflash-arxiv24,flexgen-arxiv23}
or offload data over
network~\cite{mooncake-fast25,kunserve-arxiv24,llumnix-osdi24,blitzscale-osdi25}.
Conversely, {\sys} operates at OS-level and is complementary to these
in-application swapping solutions,
as {\sys} can accurately identify the memory actually accessed at runtime.

\section{Conclusion}

This paper presents {\sys}, an OS-level scheduler tailored for
memory-oversubscribed GPU multitasking.
By leveraging predictable memory accesses,
it extends GPU context switching to include predictive working set scheduling.
Our experiments show its effectiveness on varying workloads.

\balance

\small{
\bibliographystyle{ACM-Reference-Format}
\bibliography{msched}

@inproceedings{xsched-osdi25,
  title = {{XSched}: Preemptive Scheduling for Diverse {XPU}s},
  author = {Weihang Shen and Mingcong Han and Jialong Liu and Rong Chen and Haibo Chen},
  booktitle = {19th USENIX Symposium on Operating Systems Design and Implementation (OSDI 25)},
  year = {2025},
  isbn = {978-1-939133-47-2},
  address = {Boston, MA},
  pages = {671--692},
  url = {https://www.usenix.org/conference/osdi25/presentation/shen-weihang},
  publisher = {USENIX Association},
  month = jul
}

@inproceedings {sirius-atc25,
  author={Jiali Wang and Yankui Wang and Mingcong Han and Rong Chen},
  title = {Colocating ML Inference and Training with Fast GPU Memory Handover},
  booktitle = {2025 USENIX Annual Technical Conference (USENIX ATC 25)},
  year = {2025},
  isbn = {978-1-939133-48-9},
  address = {Boston, MA},
  pages = {1657--1675},
  url = {https://www.usenix.org/conference/atc25/presentation/wang-jiali},
  publisher = {USENIX Association},
  month = jul
}

@inproceedings {reef-osdi22,
  author = {Mingcong Han and Hanze Zhang and Rong Chen and Haibo Chen},
  title = {Microsecond-scale Preemption for Concurrent {GPU-accelerated} {DNN} Inferences},
  booktitle = {16th USENIX Symposium on Operating Systems Design and Implementation (OSDI 22)},
  year = {2022},
  isbn = {978-1-939133-28-1},
  address = {Carlsbad, CA},
  pages = {539--558},
  url = {https://www.usenix.org/conference/osdi22/presentation/han},
  publisher = {USENIX Association},
  month = jul
}

@inproceedings{deepum-asplos23,
  author = {Jung, Jaehoon and Kim, Jinpyo and Lee, Jaejin},
  title = {DeepUM: Tensor Migration and Prefetching in Unified Memory},
  year = {2023},
  isbn = {9781450399166},
  publisher = {Association for Computing Machinery},
  address = {New York, NY, USA},
  url = {https://doi.org/10.1145/3575693.3575736},
  doi = {10.1145/3575693.3575736},
  booktitle = {Proceedings of the 28th ACM International Conference on Architectural Support for Programming Languages and Operating Systems, Volume 2},
  pages = {207--221},
  numpages = {15},
  location = {Vancouver, BC, Canada},
  series = {ASPLOS 2023}
}

@inproceedings{suv-micro24,
  author={B, Pratheek and Cox, Guilherme and Vesely, Jan and Basu, Arkaprava},
  booktitle={2024 57th IEEE/ACM International Symposium on Microarchitecture (MICRO)},
  title={SUV: Static Analysis Guided Unified Virtual Memory},
  year={2024},
  volume={},
  number={},
  pages={293--308},
  doi={10.1109/MICRO61859.2024.00030}
}

@misc{towards-arxiv25,
  title={Towards Efficient and Practical GPU Multitasking in the Era of LLM},
  author={Jiarong Xing and Yifan Qiao and Simon Mo and Xingqi Cui and Gur-Eyal Sela and Yang Zhou and Joseph Gonzalez and Ion Stoica},
  year={2025},
  eprint={2508.08448},
  archivePrefix={arXiv},
  primaryClass={cs.OS},
  url={https://arxiv.org/abs/2508.08448},
}

@misc{cuda-um,
  title  = {{Unified Memory Programming}},
  url    = {https://docs.nvidia.com/cuda/cuda-c-programming-guide/#unified-memory-programming},
  author = {NVIDIA},
  year   = {2025},
}

@misc{hip-um,
  title  = {{Unified memory management}},
  url    = {https://rocm.docs.amd.com/projects/HIP/en/docs-7.1.0/how-to/hip_runtime_api/memory_management/unified_memory.html},
  author = {AMD},
  year   = {2025},
}

@misc{nvlinkc2c,
  title  = {{NVIDIA NVLink-C2C}},
  url    = {https://www.nvidia.com/en-us/data-center/nvlink-c2c},
  author = {NVIDIA},
  year   = {2025},
}

@misc{cuda-um-opt,
  title  = {{Maximizing Unified Memory Performance in CUDA}},
  url    = {https://developer.nvidia.com/blog/maximizing-unified-memory-performance-cuda},
  author = {Nikolay Sakharnykh},
  year   = {2017},
}

@inproceedings{nvlinkc2c-benchmark,
  author = {Werner, Felix and Weisgut, Marcel and Rabl, Tilmann},
  title = {Towards Memory Disaggregation via NVLink C2C: Benchmarking CPU-Requested GPU Memory Access},
  year = {2025},
  isbn = {9798400714702},
  publisher = {Association for Computing Machinery},
  address = {New York, NY, USA},
  url = {https://doi.org/10.1145/3723851.3723853},
  doi = {10.1145/3723851.3723853},
  booktitle = {Proceedings of the 4th Workshop on Heterogeneous Composable and Disaggregated Systems},
  pages = {8--14},
  numpages = {7},
  location = {},
  series = {HCDS '25}
}

@inproceedings{etc-asplos19,
  author = {Li, Chen and Ausavarungnirun, Rachata and Rossbach, Christopher J. and Zhang, Youtao and Mutlu, Onur and Guo, Yang and Yang, Jun},
  title = {A Framework for Memory Oversubscription Management in Graphics Processing Units},
  year = {2019},
  isbn = {9781450362405},
  publisher = {Association for Computing Machinery},
  address = {New York, NY, USA},
  url = {https://doi.org/10.1145/3297858.3304044},
  doi = {10.1145/3297858.3304044},
  booktitle = {Proceedings of the Twenty-Fourth International Conference on Architectural Support for Programming Languages and Operating Systems},
  pages = {49--63},
  numpages = {15},
  location = {Providence, RI, USA},
  series = {ASPLOS '19}
}

@inproceedings {tgs-nsdi23,
  author = {Bingyang Wu and Zili Zhang and Zhihao Bai and Xuanzhe Liu and Xin Jin},
  title = {Transparent {GPU} Sharing in Container Clouds for Deep Learning Workloads},
  booktitle = {20th USENIX Symposium on Networked Systems Design and Implementation (NSDI 23)},
  year = {2023},
  isbn = {978-1-939133-33-5},
  address = {Boston, MA},
  pages = {69--85},
  url = {https://www.usenix.org/conference/nsdi23/presentation/wu},
  publisher = {USENIX Association},
  month = apr
}

@inproceedings {gpreempt-atc25,
  author={Ruwen Fan and Tingxu Ren and Minhui Xie and Shiwei Gao and Jiwu Shu and Youyou Lu},
  title = {{GPREEMPT: GPU Preemptive Scheduling Made General and Efficient}},
  booktitle = {2025 USENIX Annual Technical Conference (USENIX ATC 25)},
  year = {2025},
  isbn = {978-1-939133-48-9},
  address = {Boston, MA},
  pages = {263--272},
  url = {https://www.usenix.org/conference/atc25/presentation/fan},
  publisher = {USENIX Association},
  month = jul
}

@inproceedings{effisha-ppopp17,
  author = {Chen, Guoyang and Zhao, Yue and Shen, Xipeng and Zhou, Huiyang},
  title = {{EffiSha: A Software Framework for Enabling Efficient Preemptive Scheduling of GPU}},
  year = {2017},
  isbn = {9781450344937},
  publisher = {Association for Computing Machinery},
  address = {New York, NY, USA},
  url = {https://doi.org/10.1145/3018743.3018748},
  doi = {10.1145/3018743.3018748},
  booktitle = {Proceedings of the 22nd ACM SIGPLAN Symposium on Principles and Practice of Parallel Programming},
  pages = {3--16},
  numpages = {14},
  location = {Austin, Texas, USA},
  series = {PPoPP '17}
}

@inproceedings{flep-asplos17,
  author = {Wu, Bo and Liu, Xu and Zhou, Xiaobo and Jiang, Changjun},
  title = {FLEP: Enabling Flexible and Efficient Preemption on GPUs},
  year = {2017},
  isbn = {9781450344654},
  publisher = {Association for Computing Machinery},
  address = {New York, NY, USA},
  url = {https://doi.org/10.1145/3037697.3037742},
  doi = {10.1145/3037697.3037742},
  booktitle = {Proceedings of the Twenty-Second International Conference on Architectural Support for Programming Languages and Operating Systems},
  pages = {483--496},
  numpages = {14},
  location = {Xi'an, China},
  series = {ASPLOS '17}
}

@inproceedings{forest-isca25,
  author = {Lin, Mao and Feng, Yuan and Cox, Guilherme and Jeon, Hyeran},
  title = {Forest: Access-aware GPU UVM Management},
  year = {2025},
  isbn = {9798400712616},
  publisher = {Association for Computing Machinery},
  address = {New York, NY, USA},
  url = {https://doi.org/10.1145/3695053.3731047},
  doi = {10.1145/3695053.3731047},
  booktitle = {Proceedings of the 52nd Annual International Symposium on Computer Architecture},
  pages = {137--152},
  numpages = {16},
  location = {},
  series = {ISCA '25}
}

@inproceedings{earlyadaptor-ispass23,
  author={Go, Seokjin and Lee, Hyunwuk and Kim, Junsung and Lee, Jiwon and Yoon, Myung Kuk and Ro, Won Woo},
  booktitle={2023 IEEE International Symposium on Performance Analysis of Systems and Software (ISPASS)},
  title={Early-Adaptor: An Adaptive Framework for Proactive UVM Memory Management},
  year={2023},
  volume={},
  number={},
  pages={248-258},
  doi={10.1109/ISPASS57527.2023.00032}
}

@inproceedings{sentinel-hpca21,
  author={Ren, Jie and Luo, Jiaolin and Wu, Kai and Zhang, Minjia and Jeon, Hyeran and Li, Dong},
  booktitle={2021 IEEE International Symposium on High-Performance Computer Architecture (HPCA)},
  title={Sentinel: Efficient Tensor Migration and Allocation on Heterogeneous Memory Systems for Deep Learning},
  year={2021},
  volume={},
  number={},
  pages={598-611},
  doi={10.1109/HPCA51647.2021.00057}
}

@inproceedings{swapadvisor-asplos20,
  author = {Huang, Chien-Chin and Jin, Gu and Li, Jinyang},
  title = {SwapAdvisor: Pushing Deep Learning Beyond the GPU Memory Limit via Smart Swapping},
  year = {2020},
  isbn = {9781450371025},
  publisher = {Association for Computing Machinery},
  address = {New York, NY, USA},
  url = {https://doi.org/10.1145/3373376.3378530},
  doi = {10.1145/3373376.3378530},
  booktitle = {Proceedings of the Twenty-Fifth International Conference on Architectural Support for Programming Languages and Operating Systems},
  pages = {1341--1355},
  numpages = {15},
  location = {Lausanne, Switzerland},
  series = {ASPLOS '20}
}

@misc{nvidia-gsp,
  title  = {{NVIDIA RISC-V Story}},
  url    = {https://riscv.org/wp-content/uploads/2016/07/Tue1100_Nvidia_RISCV_Story_V2.pdf},
  author = {Joe Xie},
  year   = {2016},
}

@misc{nvidia-tsg,
  title  = {{NVIDIA open-gpu-doc repository}},
  url    = {https://github.com/NVIDIA/open-gpu-doc/blob/master/manuals/volta/gv100/dev_ram.ref.txt},
  author = {{NVIDIA}},
  year   = {2019}
}

@misc{amd-rr,
  title  = {{Oversubscription of hardware resources in AMD Instinct accelerators}},
  url    = {https://rocm.docs.amd.com/en/docs-6.2.2/conceptual/oversubscription.html},
  author = {{AMD}},
  year   = {2024}
}

@inproceedings{tally-asplos25,
  author = {Zhao, Wei and Jayarajan, Anand and Pekhimenko, Gennady},
  title = {Tally: Non-Intrusive Performance Isolation for Concurrent Deep Learning Workloads},
  year = {2025},
  isbn = {9798400706981},
  publisher = {Association for Computing Machinery},
  address = {New York, NY, USA},
  url = {https://doi.org/10.1145/3669940.3707282},
  doi = {10.1145/3669940.3707282},
  booktitle = {Proceedings of the 30th ACM International Conference on Architectural Support for Programming Languages and Operating Systems, Volume 1},
  pages = {1052--1068},
  numpages = {17},
  location = {Rotterdam, Netherlands},
  series = {ASPLOS '25}
}

@inproceedings{intel-gpu-context,
  author = {Choi, Wonseok and Shin, Youngjoo},
  title = {Vulnerable Intel GPU Context: Prohibit Complete Context Restore by Modifying Kernel Driver},
  year = {2025},
  isbn = {9798400714108},
  publisher = {Association for Computing Machinery},
  address = {New York, NY, USA},
  url = {https://doi.org/10.1145/3708821.3733885},
  doi = {10.1145/3708821.3733885},
  booktitle = {Proceedings of the 20th ACM Asia Conference on Computer and Communications Security},
  pages = {1632--1642},
  numpages = {11},
  location = {},
  series = {ASIA CCS '25}
}

@misc{intel-gpu-specs,
  title  = {{Intel Graphics for Linux - Programmer's Reference Manuals}},
  url    = {https://www.intel.com/content/www/us/en/docs/graphics-for-linux/developer-reference/1-0/hardware-specs.html},
  author = {Intel},
  year   = {2025},
}

@misc{nvidia-cilp,
  title  = {{Tuning CUDA Applications for Pascal: Compute Preemption}},
  url    = {https://docs.nvidia.com/cuda/pascal-tuning-guide/index.html#compute-preemption},
  author = {{NVIDIA}},
  year   = {2016},
}

@inproceedings{gcaps-ecrts24,
  author =	{Wang, Yidi and Liu, Cong and Wong, Daniel and Kim, Hyoseung},
  title =	{{GCAPS: GPU Context-Aware Preemptive Priority-Based Scheduling for Real-Time Tasks}},
  booktitle =	{36th Euromicro Conference on Real-Time Systems (ECRTS 2024)},
  pages =	{14:1--14:25},
  series =	{Leibniz International Proceedings in Informatics (LIPIcs)},
  ISBN =	{978-3-95977-324-9},
  ISSN =	{1868-8969},
  year =	{2024},
  volume =	{298},
  editor =	{Pellizzoni, Rodolfo},
  publisher =	{Schloss Dagstuhl -- Leibniz-Zentrum f{\"u}r Informatik},
  address =	{Dagstuhl, Germany},
  URL =		{https://drops.dagstuhl.de/entities/document/10.4230/LIPIcs.ECRTS.2024.14},
  URN =		{urn:nbn:de:0030-drops-203170},
  doi =		{10.4230/LIPIcs.ECRTS.2024.14}
}

@inproceedings{clockwork-osdi20,
  author = {Arpan Gujarati and Reza Karimi and Safya Alzayat and Wei Hao and Antoine Kaufmann and Ymir Vigfusson and Jonathan Mace},
  title = {Serving {DNNs} like Clockwork: Performance Predictability from the Bottom Up},
  booktitle = {14th USENIX Symposium on Operating Systems Design and Implementation (OSDI 20)},
  year = {2020},
  isbn = {978-1-939133-19-9},
  pages = {443--462},
  url = {https://www.usenix.org/conference/osdi20/presentation/gujarati},
  publisher = {USENIX Association},
  month = nov
}

@inproceedings{rammer-osdi20,
  author = {Lingxiao Ma and Zhiqiang Xie and Zhi Yang and Jilong Xue and Youshan Miao and Wei Cui and Wenxiang Hu and Fan Yang and Lintao Zhang and Lidong Zhou},
  title = {Rammer: Enabling Holistic Deep Learning Compiler Optimizations with {rTasks}},
  booktitle = {14th USENIX Symposium on Operating Systems Design and Implementation (OSDI 20)},
  year = {2020},
  isbn = {978-1-939133-19-9},
  pages = {881--897},
  url = {https://www.usenix.org/conference/osdi20/presentation/ma},
  publisher = {USENIX Association},
  month = nov
}

@article{opt-1966,
  author={Belady, L. A.},
  journal={IBM Systems Journal},
  title={A study of replacement algorithms for a virtual-storage computer},
  year={1966},
  volume={5},
  number={2},
  pages={78-101},
  doi={10.1147/sj.52.0078}
}

@inproceedings{picker-eurosys26,
  author = {Han, Mingcong and Shen, Weihang and Chen, Rong and Chen, Haibo},
  title = {Accurate and Ultra-Fast Launch-Time Validation of Idempotency for GPU Kernels},
  year = {2026},
  isbn = {9798400722127},
  publisher = {Association for Computing Machinery},
  address = {New York, NY, USA},
  url = {https://doi.org/10.1145/3767295.3803597},
  doi = {10.1145/3767295.3803597},
  booktitle = {Proceedings of the 21st European Conference on Computer Systems},
  pages = {852--971},
  numpages = {120},
  location = {McEwan Hall/The University of Edinburgh, Edinburgh, Scotland UK},
  series = {EUROSYS '26}
}

@inproceedings{phos-sosp25,
  author = {Wei, Xingda and Huang, Zhuobin and Sun, Tianle and Hao, Yingyi and Chen, Rong and Han, Mingcong and Gu, Jinyu and Chen, Haibo},
  title = {PhoenixOS: Concurrent OS-level GPU Checkpoint and Restore with Validated Speculation},
  year = {2025},
  isbn = {9798400718700},
  publisher = {Association for Computing Machinery},
  address = {New York, NY, USA},
  url = {https://doi.org/10.1145/3731569.3764813},
  doi = {10.1145/3731569.3764813},
  booktitle = {Proceedings of the ACM SIGOPS 31st Symposium on Operating Systems Principles},
  pages = {996--1013},
  numpages = {18},
  location = {Lotte Hotel World, Seoul, Republic of Korea},
  series = {SOSP '25}
}

@misc{tensorflow-malloc,
  title  = {{TensorFlow BFC Allocator}},
  url    = {https://github.com/tensorflow/tensorflow/blob/master/tensorflow/core/common_runtime/bfc_allocator.h},
  author = {{The TensorFlow Authors}},
  year   = {2015},
}

@misc{pytorch-malloc,
  title  = {{PyTorch Memory management}},
  url    = {https://docs.pytorch.org/docs/stable/notes/cuda.html#memory-management},
  author = {{PyTorch}},
  year   = {2025},
}

@misc{jax-malloc,
  title  = {{JAX GPU memory allocation}},
  url    = {https://docs.jax.dev/en/latest/gpu_memory_allocation.html},
  author = {{The JAX Authors}},
  year   = {2024},
}

@misc{llamacpp,
  title  = {{llama.cpp}},
  url    = {https://github.com/ggml-org/llama.cpp},
  author = {{ggml}},
  year   = {2025},
}

@inproceedings{nvbit-micro19,
  author = {Villa, Oreste and Stephenson, Mark and Nellans, David and Keckler, Stephen W.},
  title = {NVBit: A Dynamic Binary Instrumentation Framework for NVIDIA GPUs},
  year = {2019},
  isbn = {9781450369381},
  publisher = {Association for Computing Machinery},
  address = {New York, NY, USA},
  url = {https://doi.org/10.1145/3352460.3358307},
  doi = {10.1145/3352460.3358307},
  booktitle = {Proceedings of the 52nd Annual IEEE/ACM International Symposium on Microarchitecture},
  pages = {372--383},
  numpages = {12},
  location = {Columbus, OH, USA},
  series = {MICRO-52}
}

@book{modern-os,
  author = {Tanenbaum, Andrew S. and Bos, Herbert},
  title = {Modern Operating Systems},
  year = {2014},
  isbn = {013359162X},
  publisher = {Prentice Hall Press},
  address = {USA},
  edition = {4th}
}

@misc{linux-runqueue,
  title  = {{Linux sched.h}},
  url    = {https://github.com/torvalds/linux/blob/master/kernel/sched/sched.h},
  author = {{Linus Torvalds}},
  year   = {2025},
}

@misc{cuda-guide,
  title  = {{CUDA C++ Best Practices Guide}},
  url    = {https://docs.nvidia.com/cuda/cuda-c-best-practices-guide/},
  author = {{NVIDIA}},
  year   = {2025},
}

@misc{amd-ce,
  title  = {{System direct memory access}},
  url    = {https://rocm.docs.amd.com/en/docs-6.2.1/conceptual/gpu-memory.html#system-direct-memory-access},
  author = {{AMD}},
  year   = {2025},
}

@inproceedings{demystify-rtas24,
  author={Bakita, Joshua and Anderson, James H.},
  booktitle={2024 IEEE 30th Real-Time and Embedded Technology and Applications Symposium (RTAS)},
  title={Demystifying NVIDIA GPU Internals to Enable Reliable GPU Management},
  year={2024},
  volume={},
  number={},
  pages={294-305},
  doi={10.1109/RTAS61025.2024.00031}
}

@misc{GH200-study-arxiv24,
  title={Understanding Data Movement in Tightly Coupled Heterogeneous Systems: A Case Study with the Grace Hopper Superchip},
  author={Luigi Fusco and Mikhail Khalilov and Marcin Chrapek and Giridhar Chukkapalli and Thomas Schulthess and Torsten Hoefler},
  year={2024},
  eprint={2408.11556},
  archivePrefix={arXiv},
  primaryClass={cs.DC},
  url={https://arxiv.org/abs/2408.11556},
}

@misc{nvidia-driver-580,
  title  = {{Open GPU Kernel Modules 580.95.05}},
  url    = {https://github.com/NVIDIA/open-gpu-kernel-modules/tree/580.95.05},
  author = {{NVIDIA}},
  year   = {2025},
}

@inproceedings{rodinia-iiswc09,
  author = {Che, Shuai and Boyer, Michael and Meng, Jiayuan and Tarjan, David and Sheaffer, Jeremy W. and Lee, Sang-Ha and Skadron, Kevin},
  title = {Rodinia: A benchmark suite for heterogeneous computing},
  year = {2009},
  isbn = {9781424451562},
  publisher = {IEEE Computer Society},
  address = {USA},
  url = {https://doi.org/10.1109/IISWC.2009.5306797},
  doi = {10.1109/IISWC.2009.5306797},
  booktitle = {Proceedings of the 2009 IEEE International Symposium on Workload Characterization (IISWC)},
  pages = {44--54},
  numpages = {11},
  series = {IISWC '09}
}

@misc{arrowlake-bottleneck,
  title  = {{Intel Arrow Lake processors bottleneck PCIe 5.0 NVMe SSDs by 16\%, limiting peak speeds to 12\,GB/s instead of 14\,GB/s}},
  url    = {https://www.tomshardware.com/pc-components/cpus/intel-arrow-lake-processors-bottleneck-pcie-5-0-nvme-ssds-by-16-percent-limiting-peak-speeds-to-12gb-s-instead-of-14gb-s},
  author = {Anton Shilov and Aaron Klotz},
  year   = {2025},
}

@article{cxl-survey-csur24,
  author = {Das Sharma, Debendra and Blankenship, Robert and Berger, Daniel},
  title = {An Introduction to the Compute Express Link (CXL) Interconnect},
  year = {2024},
  issue_date = {November 2024},
  publisher = {Association for Computing Machinery},
  address = {New York, NY, USA},
  volume = {56},
  number = {11},
  issn = {0360-0300},
  url = {https://doi.org/10.1145/3669900},
  doi = {10.1145/3669900},
  journal = {ACM Comput. Surv.},
  month = jul,
  articleno = {290},
  numpages = {37}
}

@article{gpu-mt-survey-tpds22,
  author={Zhao, Chen and Gao, Wu and Nie, Feiping and Zhou, Huiyang},
  journal={IEEE Transactions on Parallel and Distributed Systems},
  title={A Survey of GPU Multitasking Methods Supported by Hardware Architecture},
  year={2022},
  volume={33},
  number={6},
  pages={1451-1463},
  doi={10.1109/TPDS.2021.3115630}
}

@inproceedings {pipeswitch-osdi20,
  author = {Zhihao Bai and Zhen Zhang and Yibo Zhu and Xin Jin},
  title = {{PipeSwitch}: Fast Pipelined Context Switching for Deep Learning Applications},
  booktitle = {14th USENIX Symposium on Operating Systems Design and Implementation (OSDI 20)},
  year = {2020},
  isbn = {978-1-939133-19-9},
  pages = {499--514},
  url = {https://www.usenix.org/conference/osdi20/presentation/bai},
  publisher = {USENIX Association},
  month = nov
}

@inproceedings {shepherd-nsdi23,
  author = {Hong Zhang and Yupeng Tang and Anurag Khandelwal and Ion Stoica},
  title = {{SHEPHERD}: Serving {DNNs} in the Wild},
  booktitle = {20th USENIX Symposium on Networked Systems Design and Implementation (NSDI 23)},
  year = {2023},
  isbn = {978-1-939133-33-5},
  address = {Boston, MA},
  pages = {787--808},
  url = {https://www.usenix.org/conference/nsdi23/presentation/zhang-hong},
  publisher = {USENIX Association},
  month = apr
}

@inproceedings{paella-sosp23,
  author = {Ng, Kelvin K. W. and Demoulin, Henri Maxime and Liu, Vincent},
  title = {Paella: Low-latency Model Serving with Software-defined GPU Scheduling},
  year = {2023},
  isbn = {9798400702297},
  publisher = {Association for Computing Machinery},
  address = {New York, NY, USA},
  url = {https://doi.org/10.1145/3600006.3613163},
  doi = {10.1145/3600006.3613163},
  booktitle = {Proceedings of the 29th Symposium on Operating Systems Principles},
  pages = {595--610},
  numpages = {16},
  location = {Koblenz, Germany},
  series = {SOSP '23}
}

@inproceedings{lithos-sosp25,
  author = {Coppock, Patrick H. and Zhang, Brian and Solomon, Eliot H. and Kypriotis, Vasilis and Yang, Leon and Sharma, Bikash and Schatzberg, Dan and Mowry, Todd C. and Skarlatos, Dimitrios},
  title = {LithOS: An Operating System for Efficient Machine Learning on GPUs},
  year = {2025},
  isbn = {9798400718700},
  publisher = {Association for Computing Machinery},
  address = {New York, NY, USA},
  url = {https://doi.org/10.1145/3731569.3764818},
  doi = {10.1145/3731569.3764818},
  booktitle = {Proceedings of the ACM SIGOPS 31st Symposium on Operating Systems Principles},
  pages = {1--17},
  numpages = {17},
  location = {Lotte Hotel World, Seoul, Republic of Korea},
  series = {SOSP '25}
}

@inproceedings{chimera-asplos15,
  author = {Park, Jason Jong Kyu and Park, Yongjun and Mahlke, Scott},
  title = {Chimera: Collaborative Preemption for Multitasking on a Shared GPU},
  year = {2015},
  isbn = {9781450328357},
  publisher = {Association for Computing Machinery},
  address = {New York, NY, USA},
  url = {https://doi.org/10.1145/2694344.2694346},
  doi = {10.1145/2694344.2694346},
  booktitle = {Proceedings of the Twentieth International Conference on Architectural Support for Programming Languages and Operating Systems},
  pages = {593--606},
  numpages = {14},
  location = {Istanbul, Turkey},
  series = {ASPLOS '15}
}

@misc{nvidia-mps,
  title  = {{Multi-Process Service}},
  url    = {https://docs.nvidia.com/deploy/mps/index.html},
  author = {NVIDIA},
  year   = {2025},
}

@misc{nvidia-mig,
  title  = {{MIG User Guide}},
  url    = {https://docs.nvidia.com/datacenter/tesla/mig-user-guide/index.html},
  author = {NVIDIA},
  year   = {2025},
}

@inproceedings{libsmctrl-rtas23,
  author={Bakita, Joshua and Anderson, James H.},
  booktitle={2023 IEEE 29th Real-Time and Embedded Technology and Applications Symposium (RTAS)},
  title={Hardware Compute Partitioning on NVIDIA GPUs},
  year={2023},
  volume={},
  number={},
  pages={54-66},
  doi={10.1109/RTAS58335.2023.00012}
}

@inproceedings{baymax-asplos16,
  author = {Chen, Quan and Yang, Hailong and Mars, Jason and Tang, Lingjia},
  title = {Baymax: QoS Awareness and Increased Utilization for Non-Preemptive Accelerators in Warehouse Scale Computers},
  year = {2016},
  isbn = {9781450340915},
  publisher = {Association for Computing Machinery},
  address = {New York, NY, USA},
  url = {https://doi.org/10.1145/2872362.2872368},
  doi = {10.1145/2872362.2872368},
  booktitle = {Proceedings of the Twenty-First International Conference on Architectural Support for Programming Languages and Operating Systems},
  pages = {681--696},
  numpages = {16},
  location = {Atlanta, Georgia, USA},
  series = {ASPLOS '16}
}

@inproceedings{orion-eurosys24,
  author = {Strati, Foteini and Ma, Xianzhe and Klimovic, Ana},
  title = {Orion: Interference-aware, Fine-grained GPU Sharing for ML Applications},
  year = {2024},
  isbn = {9798400704376},
  publisher = {Association for Computing Machinery},
  address = {New York, NY, USA},
  url = {https://doi.org/10.1145/3627703.3629578},
  doi = {10.1145/3627703.3629578},
  booktitle = {Proceedings of the Nineteenth European Conference on Computer Systems},
  pages = {1075--1092},
  numpages = {18},
  location = {Athens, Greece},
  series = {EuroSys '24}
}

@inproceedings{sgdrc-ppopp25,
  author = {Zhang, Yongkang and Yu, Haoxuan and Han, Chenxia and Wang, Cheng and Lu, Baotong and Li, Yunzhe and Jiang, Zhifeng and Li, Yang and Chu, Xiaowen and Li, Huaicheng},
  title = {SGDRC: Software-Defined Dynamic Resource Control for Concurrent DNN Inference on NVIDIA GPUs},
  year = {2025},
  isbn = {9798400714436},
  publisher = {Association for Computing Machinery},
  address = {New York, NY, USA},
  url = {https://doi.org/10.1145/3710848.3710863},
  doi = {10.1145/3710848.3710863},
  booktitle = {Proceedings of the 30th ACM SIGPLAN Annual Symposium on Principles and Practice of Parallel Programming},
  pages = {267--281},
  numpages = {15},
  location = {Las Vegas, NV, USA},
  series = {PPoPP '25}
}

@inproceedings{snurhac-hpdc21,
  author = {Jung, Jaehoon and Park, Daeyoung and Jo, Gangwon and Park, Jungho and Lee, Jaejin},
  title = {SnuRHAC: A Runtime for Heterogeneous Accelerator Clusters with CUDA Unified Memory},
  year = {2021},
  isbn = {9781450382175},
  publisher = {Association for Computing Machinery},
  address = {New York, NY, USA},
  url = {https://doi.org/10.1145/3431379.3460647},
  doi = {10.1145/3431379.3460647},
  booktitle = {Proceedings of the 30th International Symposium on High-Performance Parallel and Distributed Computing},
  pages = {107--120},
  numpages = {14},
  location = {Virtual Event, Sweden},
  series = {HPDC '21}
}

@inproceedings{huvm-atc22,
  author = {Sangjin Choi and Taeksoo Kim and Jinwoo Jeong and Rachata Ausavarungnirun and Myeongjae Jeon and Youngjin Kwon and Jeongseob Ahn},
  title = {Memory Harvesting in {Multi-GPU} Systems with Hierarchical Unified Virtual Memory},
  booktitle = {2022 USENIX Annual Technical Conference (USENIX ATC 22)},
  year = {2022},
  isbn = {978-1-939133-29-66},
  address = {Carlsbad, CA},
  pages = {625--638},
  url = {https://www.usenix.org/conference/atc22/presentation/choi-sangjin},
  publisher = {USENIX Association},
  month = jul
}

@inproceedings{smc-icpp24,
  author = {Nihaal, Abdun and Mutyam, Madhu},
  title = {Selective Memory Compression for GPU Memory Oversubscription Management},
  year = {2024},
  isbn = {9798400717932},
  publisher = {Association for Computing Machinery},
  address = {New York, NY, USA},
  url = {https://doi.org/10.1145/3673038.3673058},
  doi = {10.1145/3673038.3673058},
  booktitle = {Proceedings of the 53rd International Conference on Parallel Processing},
  pages = {189--198},
  numpages = {10},
  location = {Gotland, Sweden},
  series = {ICPP '24}
}

@inproceedings{capuchin-asplos20,
  author = {Peng, Xuan and Shi, Xuanhua and Dai, Hulin and Jin, Hai and Ma, Weiliang and Xiong, Qian and Yang, Fan and Qian, Xuehai},
  title = {Capuchin: Tensor-based GPU Memory Management for Deep Learning},
  year = {2020},
  isbn = {9781450371025},
  publisher = {Association for Computing Machinery},
  address = {New York, NY, USA},
  url = {https://doi.org/10.1145/3373376.3378505},
  doi = {10.1145/3373376.3378505},
  booktitle = {Proceedings of the Twenty-Fifth International Conference on Architectural Support for Programming Languages and Operating Systems},
  pages = {891--905},
  numpages = {15},
  location = {Lausanne, Switzerland},
  series = {ASPLOS '20}
}

@inproceedings{powerinfer-sosp24,
  author = {Song, Yixin and Mi, Zeyu and Xie, Haotong and Chen, Haibo},
  title = {PowerInfer: Fast Large Language Model Serving with a Consumer-grade GPU},
  year = {2024},
  isbn = {9798400712517},
  publisher = {Association for Computing Machinery},
  address = {New York, NY, USA},
  url = {https://doi.org/10.1145/3694715.3695964},
  doi = {10.1145/3694715.3695964},
  booktitle = {Proceedings of the ACM SIGOPS 30th Symposium on Operating Systems Principles},
  pages = {590--606},
  numpages = {17},
  location = {Austin, TX, USA},
  series = {SOSP '24}
}

@inproceedings{pagedattention-sosp23,
  author = {Kwon, Woosuk and Li, Zhuohan and Zhuang, Siyuan and Sheng, Ying and Zheng, Lianmin and Yu, Cody Hao and Gonzalez, Joseph and Zhang, Hao and Stoica, Ion},
  title = {Efficient Memory Management for Large Language Model Serving with PagedAttention},
  year = {2023},
  isbn = {9798400702297},
  publisher = {Association for Computing Machinery},
  address = {New York, NY, USA},
  url = {https://doi.org/10.1145/3600006.3613165},
  doi = {10.1145/3600006.3613165},
  booktitle = {Proceedings of the 29th Symposium on Operating Systems Principles},
  pages = {611--626},
  numpages = {16},
  location = {Koblenz, Germany},
  series = {SOSP '23}
}

@misc{lmcache-arxiv25,
  title={LMCache: An Efficient KV Cache Layer for Enterprise-Scale LLM Inference},
  author={Yuhan Liu and Yihua Cheng and Jiayi Yao and Yuwei An and Xiaokun Chen and Shaoting Feng and Yuyang Huang and Samuel Shen and Rui Zhang and Kuntai Du and Junchen Jiang},
  year={2025},
  eprint={2510.09665},
  archivePrefix={arXiv},
  primaryClass={cs.LG},
  url={https://arxiv.org/abs/2510.09665},
}

@misc{flexgen-arxiv23,
  title={FlexGen: High-Throughput Generative Inference of Large Language Models with a Single GPU},
  author={Ying Sheng and Lianmin Zheng and Binhang Yuan and Zhuohan Li and Max Ryabinin and Daniel Y. Fu and Zhiqiang Xie and Beidi Chen and Clark Barrett and Joseph E. Gonzalez and Percy Liang and Christopher Ré and Ion Stoica and Ce Zhang},
  year={2023},
  eprint={2303.06865},
  archivePrefix={arXiv},
  primaryClass={cs.LG},
  url={https://arxiv.org/abs/2303.06865},
}

@inproceedings{moelightning-asplos25,
  author = {Cao, Shiyi and Liu, Shu and Griggs, Tyler and Schafhalter, Peter and Liu, Xiaoxuan and Sheng, Ying and Gonzalez, Joseph E. and Zaharia, Matei and Stoica, Ion},
  title = {MoE-Lightning: High-Throughput MoE Inference on Memory-constrained GPUs},
  year = {2025},
  isbn = {9798400706981},
  publisher = {Association for Computing Machinery},
  address = {New York, NY, USA},
  url = {https://doi.org/10.1145/3669940.3707267},
  doi = {10.1145/3669940.3707267},
  booktitle = {Proceedings of the 30th ACM International Conference on Architectural Support for Programming Languages and Operating Systems, Volume 1},
  pages = {715--730},
  numpages = {16},
  location = {Rotterdam, Netherlands},
  series = {ASPLOS '25}
}

@misc{llminflash-arxiv24,
  title={LLM in a flash: Efficient Large Language Model Inference with Limited Memory},
  author={Keivan Alizadeh and Iman Mirzadeh and Dmitry Belenko and Karen Khatamifard and Minsik Cho and Carlo C Del Mundo and Mohammad Rastegari and Mehrdad Farajtabar},
  year={2024},
  eprint={2312.11514},
  archivePrefix={arXiv},
  primaryClass={cs.CL},
  url={https://arxiv.org/abs/2312.11514},
}

@inproceedings{mooncake-fast25,
  author = {Ruoyu Qin and Zheming Li and Weiran He and Jialei Cui and Feng Ren and Mingxing Zhang and Yongwei Wu and Weimin Zheng and Xinran Xu},
  title = {Mooncake: Trading More Storage for Less Computation {\textemdash} A {KVCache-centric} Architecture for Serving {LLM} Chatbot},
  booktitle = {23rd USENIX Conference on File and Storage Technologies (FAST 25)},
  year = {2025},
  isbn = {978-1-939133-45-8},
  address = {Santa Clara, CA},
  pages = {155--170},
  url = {https://www.usenix.org/conference/fast25/presentation/qin},
  publisher = {USENIX Association},
  month = feb
}

@misc{kunserve-arxiv24,
  title={KunServe: Parameter-centric Memory Management for Efficient Memory Overloading Handling in LLM Serving},
  author={Rongxin Cheng and Yuxin Lai and Xingda Wei and Rong Chen and Haibo Chen},
  year={2025},
  eprint={2412.18169},
  archivePrefix={arXiv},
  primaryClass={cs.DC},
  url={https://arxiv.org/abs/2412.18169},
}

@inproceedings{llumnix-osdi24,
  author = {Biao Sun and Ziming Huang and Hanyu Zhao and Wencong Xiao and Xinyi Zhang and Yong Li and Wei Lin},
  title = {Llumnix: Dynamic Scheduling for Large Language Model Serving},
  booktitle = {18th USENIX Symposium on Operating Systems Design and Implementation (OSDI 24)},
  year = {2024},
  isbn = {978-1-939133-40-3},
  address = {Santa Clara, CA},
  pages = {173--191},
  url = {https://www.usenix.org/conference/osdi24/presentation/sun-biao},
  publisher = {USENIX Association},
  month = jul
}

@inproceedings{blitzscale-osdi25,
  author = {Zhang, Dingyan and Wang, Haotian and Liu, Yang and Wei, Xingda and Shan, Yizhou and Chen, Rong and Chen, Haibo},
  title = {BLITZSCALE: fast and live large model autoscaling with O(1) host caching},
  year = {2025},
  isbn = {978-1-939133-47-2},
  publisher = {USENIX Association},
  address = {USA},
  booktitle = {Proceedings of the 19th USENIX Conference on Operating Systems Design and Implementation},
  articleno = {16},
  numpages = {19},
  location = {Boston, MA, USA},
  series = {OSDI '25}
}

@misc{powerinfer2-arxiv24,
  title={PowerInfer-2: Fast Large Language Model Inference on a Smartphone},
  author={Zhenliang Xue and Yixin Song and Zeyu Mi and Xinrui Zheng and Yubin Xia and Haibo Chen},
  year={2024},
  eprint={2406.06282},
  archivePrefix={arXiv},
  primaryClass={cs.LG},
  url={https://arxiv.org/abs/2406.06282},
}

@inproceedings{aegaeon-sosp25,
  author = {Xiang, Yuxing and Li, Xue and Qian, Kun and Yang, Yufan and Zhu, Diwen and Yu, Wenyuan and Zhai, Ennan and Liu, Xuanzhe and Jin, Xin and Zhou, Jingren},
  title = {Aegaeon: Effective GPU Pooling for Concurrent LLM Serving on the Market},
  year = {2025},
  isbn = {9798400718700},
  publisher = {Association for Computing Machinery},
  address = {New York, NY, USA},
  url = {https://doi.org/10.1145/3731569.3764815},
  doi = {10.1145/3731569.3764815},
  booktitle = {Proceedings of the ACM SIGOPS 31st Symposium on Operating Systems Principles},
  pages = {1030--1045},
  numpages = {16},
  location = {Lotte Hotel World, Seoul, Republic of Korea},
  series = {SOSP '25}
}

@article{multimedia-tcc20,
  author={Li, He and Ota, Kaoru and Dong, Mianxiong and Vasilakos, Athanasios V. and Nagano, Koji},
  journal={IEEE Transactions on Cloud Computing},
  title={Multimedia Processing Pricing Strategy in GPU-Accelerated Cloud Computing},
  year={2020},
  volume={8},
  number={4},
  pages={1264-1273},
  doi={10.1109/TCC.2017.2672554}
}

@article{clij-naturemethods20,
	title = {{CLIJ}: {GPU}-accelerated image processing for everyone},
	volume = {17},
	issn = {1548-7105},
	url = {https://doi.org/10.1038/s41592-019-0650-1},
	doi = {10.1038/s41592-019-0650-1},
	number = {1},
	journal = {Nature Methods},
	author = {Haase, Robert and Royer, Loic A. and Steinbach, Peter and Schmidt, Deborah and Dibrov, Alexandr and Schmidt, Uwe and Weigert, Martin and Maghelli, Nicola and Tomancak, Pavel and Jug, Florian and Myers, Eugene W.},
	month = jan,
	year = {2020},
	pages = {5--6},
}

@inproceedings{vectorsearch-fast25,
  author = {Bing Tian and Haikun Liu and Yuhang Tang and Shihai Xiao and Zhuohui Duan and Xiaofei Liao and Hai Jin and Xuecang Zhang and Junhua Zhu and Yu Zhang},
  title = {Towards High-throughput and Low-latency Billion-scale Vector Search via {CPU/GPU} Collaborative Filtering and Re-ranking},
  booktitle = {23rd USENIX Conference on File and Storage Technologies (FAST 25)},
  year = {2025},
  isbn = {978-1-939133-45-8},
  address = {Santa Clara, CA},
  pages = {171--185},
  url = {https://www.usenix.org/conference/fast25/presentation/tian-bing},
  publisher = {USENIX Association},
  month = feb
}

@article{gpudb-vldb23,
  author = {Cao, Jiashen and Sen, Rathijit and Interlandi, Matteo and Arulraj, Joy and Kim, Hyesoon},
  title = {GPU Database Systems Characterization and Optimization},
  year = {2023},
  issue_date = {November 2023},
  publisher = {VLDB Endowment},
  volume = {17},
  number = {3},
  issn = {2150-8097},
  url = {https://doi.org/10.14778/3632093.3632107},
  doi = {10.14778/3632093.3632107},
  journal = {Proc. VLDB Endow.},
  month = nov,
  pages = {441--454},
  numpages = {14}
}

@inproceedings{aiservice-apsys25,
  author = {Yang, Jinrong and Wang, Zimeng and Chen, Rong and Chen, Haibo},
  title = {A System-level Abstraction and Service for Flourishing AI-powered Applications},
  year = {2025},
  isbn = {9798400715723},
  publisher = {Association for Computing Machinery},
  address = {New York, NY, USA},
  url = {https://doi.org/10.1145/3725783.3764406},
  doi = {10.1145/3725783.3764406},
  booktitle = {Proceedings of the 16th ACM SIGOPS Asia-Pacific Workshop on Systems},
  pages = {106--114},
  numpages = {9},
  location = {Lotte Hotel World, Emerald Hall, Seoul, Republic of Korea},
  series = {APSys '25}
}

@misc{smartapp-arxiv23,
  title={Towards Real Smart Apps: Investigating Human-AI Interactions in Smartphone On-Device AI Apps},
  author={Jason Ching Yuen Siu and Jieshan Chen and Yujin Huang and Zhenchang Xing and Chunyang Chen},
  year={2023},
  eprint={2307.00756},
  archivePrefix={arXiv},
  primaryClass={cs.HC},
  url={https://arxiv.org/abs/2307.00756},
}

@misc{aiui-arxiv24,
  title={Generative AI in Multimodal User Interfaces: Trends, Challenges, and Cross-Platform Adaptability},
  author={J. Bieniek and M. Rahouti and D. C. Verma},
  year={2024},
  eprint={2411.10234},
  archivePrefix={arXiv},
  primaryClass={cs.HC},
  url={https://arxiv.org/abs/2411.10234},
}

@misc{crewai-github,
  title  = {{crewAI}},
  url    = {https://github.com/crewAIInc/crewAI},
  author = {{crewAIInc}},
  year   = {2026},
}

@misc{opencode-github,
  title  = {{opencode}},
  url    = {https://github.com/anomalyco/opencode},
  author = {{anomalyco}},
  year   = {2026},
}

@misc{anthropic-harness,
  title  = {Harness design for long-running application development},
  url    = {https://www.anthropic.com/engineering/harness-design-long-running-apps},
  author = {Prithvi Rajasekaran},
  year   = {2026},
}

@misc{superpowers-parallel,
  title  = {Dispatching Parallel Agents},
  url    = {https://github.com/obra/superpowers/blob/main/skills/dispatching-parallel-agents/SKILL.md},
  author = {Jesse Vincent},
  year   = {2026},
}

@inproceedings{nvshare-icse24,
  author={Alexopoulos, Georgios and Mitropoulos, Dimitris},
  booktitle={2024 IEEE/ACM 46th International Conference on Software Engineering: Companion Proceedings (ICSE-Companion)},
  title={nvshare: Practical GPU Sharing Without Memory Size Constraints},
  year={2024},
  volume={},
  number={},
  pages={16--20},
  doi={10.1145/3639478.3640034}
}

@misc{vllm-github,
  title  = {{vLLM}},
  url    = {https://github.com/vllm-project/vllm},
  author = {vllm-project},
  year   = {2026},
}

@misc{cupti-nvidia,
  title  = {{CUPTI}},
  url    = {https://docs.nvidia.com/cupti/index.html},
  author = {{NVIDIA}},
  year   = {2026},
}

@misc{vllm-sleep,
  title  = {Zero-Reload Model Switching with vLLM Sleep Mode},
  url    = {https://vllm.ai/blog/sleep-mode},
  author = {{vLLM}},
  year   = {2026},
}

@inproceedings{moe-predict-isca26,
  author={Yu, Zhongkai and Guan, Yue and Yu, Zihao and Zhou, Chenyang and Hu, Zhengding and Pei, Shuyi and Kang, Yangwook and Ding, Yufei and Tsai, Po-An},
  booktitle={2026 ACM/IEEE 53rd Annual International Symposium on Computer Architecture (ISCA)},
  title={Patterns Behind Chaos: Forecasting Data Movement for Efficient Large-Scale Moe LLM Inference},
  year={2026},
  volume={},
  number={},
  pages={68--84},
  doi={10.1109/ISCA66397.2026.00021}
}

@inproceedings{klotski-asplos25,
  author = {Fang, Zhiyuan and Huang, Yuegui and Hong, Zicong and Lyu, Yufeng and Chen, Wuhui and Yu, Yue and Yu, Fan and Zheng, Zibin},
  title = {Klotski: Efficient Mixture-of-Expert Inference via Expert-Aware Multi-Batch Pipeline},
  year = {2025},
  isbn = {9798400710797},
  publisher = {Association for Computing Machinery},
  address = {New York, NY, USA},
  url = {https://doi.org/10.1145/3676641.3716261},
  doi = {10.1145/3676641.3716261},
  booktitle = {Proceedings of the 30th ACM International Conference on Architectural Support for Programming Languages and Operating Systems, Volume 2},
  pages = {574--588},
  numpages = {15},
  location = {Rotterdam, Netherlands},
  series = {ASPLOS '25}
}

@inproceedings{lina-atc23,
  author = {Jiamin Li and Yimin Jiang and Yibo Zhu and Cong Wang and Hong Xu},
  title = {Accelerating Distributed {MoE} Training and Inference with Lina},
  booktitle = {2023 USENIX Annual Technical Conference (USENIX ATC 23)},
  year = {2023},
  isbn = {978-1-939133-35-9},
  address = {Boston, MA},
  pages = {945--959},
  url = {https://www.usenix.org/conference/atc23/presentation/li-jiamin},
  publisher = {USENIX Association},
  month = jul
}

@inproceedings{nixie-osdi26,
  author = {Yechen Xu and Yifei Wang and Nathanael Ren and Yiran Chen and Danyang Zhuo},
  title = {Nixie: Efficient, Transparent Temporal Multiplexing for Consumer {GPUs}},
  booktitle = {20th USENIX Symposium on Operating Systems Design and Implementation (OSDI 26)},
  year = {2026},
  isbn = {978-1-939133-55-7},
  address = {Seattle, WA},
  pages = {2085--2101},
  url = {https://www.usenix.org/conference/osdi26/presentation/xu-yechen},
  publisher = {USENIX Association},
  month = jul
}

@article{gvm-paper,
  title={Breaking the Tradeoff: Elastic and Isolated GPU Sharing with GVM},
  author={Liu, Yicheng and Qiao, Yifan and Xia, Tian and Xu, Yi and Hong, Tony and Zhou, Yutong and Sun, David and Yang, Shuo and Zhao, Yilong and Shu, Junyi and others},
  year = {2026},
  url={https://web.cs.ucla.edu/~harryxu/papers/ghost-asplos27.pdf}
}

@misc{gvm-github,
  title  = {{GVM}},
  url    = {https://github.com/ovg-project/GVM},
  author = {{ovg-project}},
  year   = {2026},
}

@misc{sharegpt-huggingface,
  title  = {{ShareGPT} {Dataset}},
  url    = {https://huggingface.co/datasets/anon8231489123/ShareGPT_Vicuna_unfiltered},
  author = {{anon8231489123}},
  year   = {2023},
}
}

\end{document}